**И. З. ШКУРЧЕНКО**

# СТРОЕНИЕ СОЛНЦА И ПЛАНЕТ СОЛНЕЧНОЙ СИСТЕМЫ С ТОЧКИ ЗРЕНИЯ МЕХАНИКИ БЕЗЫНЕРТНОЙ МАССЫ II

Часть II является продолжением и завершением монографии «Строение Солнца и планет солнечной системы с точки зрения механики безынертной массы массы I». Вторая часть содержит исследование строения Солнца и планет. Монография адресована специалистам в области теоретической и практической гидродинамики и смежных наук. Она будет полезной для астрономов, синоптиков и геологов.

**I. Z. SHKURCHENKO**

# THE STRUCTURE OF THE SUN AND THE PLANETS OF THE SOLAR SYSTEM FROM THE VIEWPOINT OF MECHANICS OF THE INERTLESS MASS II

The second part of the monograph contains the investigation of the structure of the Sun and planets of the solar system. This monograph is addressed to specialists in the field of theoretical and practical hydrodynamics and adjacent sciences. It will be useful for astronomers, meteorologists and geologists.

**СОДЕРЖАНИЕ ЧАСТИ II:**

## Глава 5. СОЛНЦЕ И ПЛАНЕТЫ СОЛНЕЧНОЙ СИСТЕМЫ[1]

В начале нашего путешествия мы ознакомимся с тем, что уже известно о Солнце и планетах солнечной системы. Познакомит нас с ними учебник для десятого класса средней школы за 1973 год, автором которого является Б.А. Воронцов-Вельяминов.

**Солнце.** Принято считать, что Солнце является самосветящимся газовым шаром, черпающим из своих недр колоссальные запасы энергии. Источником энергии являются ядерные реакции. Солнце относится к разряду жёлтых звезд. Видимую поверхность Солнца называют фотосферой. Толщина её, как слоя, около 300 км. Плотность этой солнечной зоны не велика, а давление составляет около 0,1 земной атмосферы.

Над фотосферой находится атмосфера Солнца. Минимальная температура фотосферы около 4400°К. Нижняя часть атмосферы называется хромосферой. В хромосфере температура постоянно растет до нескольких десятков тысяч градусов. Хромосфера гораздо разреженнее, чем фотосфера. На фоне яркого неба её не видно. Выше хромосферы над Солнцем простирается самая верхняя часть его атмосферы – солнечная корона. Она состоит из разреженного газа, имеющего температуру около миллиона градусов. Солнечную корону мы наблюдаем в виде яркого веера лучей, расположенного по окружности Солнца. В атмосфере Солнца найдено 68 элементов Периодической таблицы Д. И. Менделеева. Считают, что массу Солнца на 70% составляет водород, 29% – гелий и на все остальные элементы приходится 1%. В составе Солнца находятся те же элементы, которые имеются на Земле.

Солнце имеет идеальную форму шара, диаметр которого равен 1392000 км. Он в 109,1 раз превышает диаметр Земли. Масса Солнца больше массы всех планет. Средняя плотность Солнца равна 1,4 г/см$^3$. Солнце вращается вокруг собственной оси зонами – быстрее всего на экваторе, где период вращения составляет 25 суток. К полюсу период увеличивается до 30 суток. Солнце имеет магнитное и гравитационное силовые поля.

**Земля.** Земля как планета обладает формой, близкой к шаровой поверхности. Экваториальный радиус Земли на 21,4 км больше полярного. Экваториальный радиус Земли, по данным советских ученых, равен 6 378,2 км. Когда Землю для простоты принимают за шар, то за его радиус берут 6371 км – радиус шара, равновеликого Земле. Поверхность её образуют твёрдые минералы. Толщина твёрдой поверхности Земли доходит до нескольких десятков километров. Под твёрдой поверхностью Земли находятся расплавленные минералы, которые имеют температуру в несколько тысяч градусов. При этой температуре они находятся в виде жидкости с определённой плотностью, которую называют магмой. Где-то в центре Земли, как полагают, существует ядро Земли какого-то иного структурного построения.

Выше твёрдой поверхности Земли находится её атмосфера. Она содержит по объёму 78% азота, 21% кислорода и ничтожное количество других газов, в том числе и водяные пары. Нижний слой атмосферы называют тропосферой, которая простирается в средних широтах до высоты 10 – 12 км. В ней температура падает с высотой. Затем идёт следующий слой атмосферы, который называют стратосферой, где температура постоянна и равна в среднем –40°С. С высот около 25 км температура атмосферы медленно растет. Верхнюю границу стратосферы принимают на высоте порядка 80 км. Плотность атмосферы падает с высотой. На высоте 100 км давление в миллион раз меньше, чем на уровне моря. Выше стратосферы располагается ионосфера, границы которой простираются до высоты порядка 1800 км. Считают, что ионосферу составляют очень разреженные, ионизированные Солнцем газы. Здесь и до высоты в несколько радиусов Земли имеется только разреженный водород, образующий геокорону. Её плотность порядка сотен атомов в кубическом сантиметре. По этой причине считают, что атмосфера Земли через эти слои постепенно рассеивается.

Один оборот вокруг собственной оси Земля совершает за 23 часа 56 минут и 4 секунды. Средняя плотность Земли равна 5,5 г/см$^3$. У Земли есть гравитационное и магнитное силовые поля. Земля имеет свой собственный спутник, который называют Луной. Луна по диаметру вчетверо, а по массе в 81 раз меньше Земли. Средняя плотность Луны 3,3 г/см$^3$. Она не имеет своей атмосферы. Луна вращается вокруг Земли в направлении её осевого вращения. Солнечные сутки Луны равны 29,5 земным суткам.

К планетам земной группы относятся также планеты Меркурий, Венера и Марс.

**Меркурий.** Самая близкая к Солнцу планета, немногим больше Луны. Диаметр его равен 4900 км. Меркурий имеет форму идеальной шаровой поверхности без всякого сжатия. Средняя плотность равна 5,6 г/см$^3$. Один оборот вокруг собственной оси он делает за 58,65 земных суток. Спутников не имеет. По ряду признаков полагают о возможном существовании у Меркурия атмосферы. Температура на поверхности Меркурия равна 300°С.

[1] В продолжение темы строения небесных тел, см. работу «Энергетическая структура Солнца и планет солнечной системы с точки зрения механики безынертной массы» (1977 г.).

**Венера.** Эта планета интересна тем, что она по объёму и по массе только немного меньше Земли. Экваториальный диаметр её равен 12100 км. Общая форма Венеры имеет идеальную шаровую поверхность. Один оборот вокруг своей оси Венера делает за 243,0 суток. Она вращается вокруг оси в сторону, противоположную той, в которую вращаются все планеты (кроме Урана) и в которую она сама обращается вокруг Солнца.

Венера имеет атмосферу. По данным автоматических станций «Венера – 4», «5», «6», «7» и «Маринера – 5», которые мы берём из журнала «Наука и жизнь», № 4 за 1972 год, она имеет следующее строение:

слой атмосферы от поверхности Венеры до высоты 50 – 60 км называют тропосферой. Стратосфера располагается над тропосферой до высоты 100 – 110 км. Выше располагается ионосфера. На высоте 60 – 70 км имеется облака из $H_2O$, которые состоят из ледяных кристалликов. Толщина слоя облаков – 7–10 км.

Полагают такой состав атмосферы Венеры: около 95% углекислого газа, около 4% азота и менее 0,4% кислорода. Температура на среднем уровне твёрдой поверхности Венеры около 500°С, а давление около 100 атмосфер. По данным учебника «Астрономия», измерения температуры надоблачного слоя планеты дали около –40°С и на дневном, и на ночном полушарии. Автоматические станции, пролетевшие мимо Венеры, не обнаружили у неё магнитного поля и радиационных поясов. Упустили мы записать среднюю плотность Венеры, которая равна 5,2 г/см$^3$.

**Марс.** По размеру планета занимает промежуточное положение между Землей и Луной. Экваториальный диаметр его равен 6800 км. Полярный радиус меньше экваториального на 23 км. Средняя его плотность равна 4,0 г/см$^3$, а общая масса Марса составляет от Земной массы всего 0,53. Один оборот вокруг своей оси вращения он делает за 24 часа 37 минут и 23 секунды. У Марса имеются два небольших спутника. Считают, что атмосфера Марса содержит не более 0,001 кислорода от его содержания в земной атмосфере. Давление атмосферы на Марсе составляет лишь около 0,006 давления земной атмосферы. Наличие полярных шапок и белых облаков говорит о присутствии воды на Марсе в сравнительно небольших количествах. Температура поверхности на экваторе достигает 10 – 30°С. К вечеру она падает до –60°С. Магнитное поле и радиационный пояс у Марса не обнаружены.

**Планеты - гиганты.** К этой группе планет относят Юпитер, Сатурн, Уран и Нептун.

Из этих планет лучше всех изучен Юпитер – самая большая и ближайшая из этой группы к Солнцу планета. Он в 11 раз больше Земли по диаметру и в 318 раз – по массе. Экваториальный диаметр Юпитера равен 142000 км. Полярный радиус его на 4 437 км меньше экваториального. Средняя плотность Юпитера – 1,3 г/см$^3$.

Сатурн, экваториальный диаметр которого равен 120000 км, имеет среднюю плотность 0,7 г/см$^3$. Полярный радиус его на 6000 км меньше экваториального. Уран имеет среднюю плотность, равную 1,5 г/см$^3$. Экваториальный диаметр его равен 50000 км. Полярный радиус его на 625 км меньше экваториального. Нептун, средняя плотность которого равна 1,7 г/см$^3$, имеет экваториальный диаметр равный 50000 км. Полярный радиус меньше экваториального на 417 км.

Все планеты - гиганты окружены мощными протяжёнными атмосферами. Тёмные облака, по-видимому, выше светлых, и яснее всего они видны на Юпитере. Никаких постоянных деталей на Юпитере нет, кроме красного пятна. Температура этих планет (по крайней мере, над облаками) очень низка: на Юпитере –145°С, на Сатурне –180°С, на Уране и Нептуне ещё ниже. Спектральные наблюдения показывают, что атмосферы планет - гигантов содержат в основном молекулярный водород и метан $CH_4$, а в атмосфере Юпитера есть ещё и аммиак $NH_3$. Юпитер вращается зонами: чем ближе к полюсу, тем медленнее. На экваторе период вращения – 9 час. 50 мин, а у красного пятна на 5 мин 11 сек больше. Период вращения на экваторе Сатурна – 10 час. 14 мин, Урана – 19 час. 49 мин, Нептуна – 15 час. 49 мин.

Сатурн имеет плоское кольцо, толщиной в несколько километров. Оно расположено в плоскости экватора планеты. Ширина этого кольца такова, что по нему мог бы катиться Земной шар. Считают, что кольцо Сатурна составляют частицы с размерами от 1 см до 1 м. Особенность Урана заключается в том, что ось его образует с плоскостью орбиты угол всего лишь 8°, так что он вращается как бы лежа на боку. Уран и Венера – единственные планеты, вращающиеся не в ту сторону, в которую вращаются все остальные.

Больше всего спутников у Юпитера, которых насчитывается 12. Сатурн имеет 10 спутников, у Урана – 5, а у Нептуна – 2. Самые крупные из спутников: Титан (спутник Сатурна) и Ганимед (третий спутник Юпитера). Они в полтора раза больше Луны по диаметру и немного больше Меркурия. Титан – единственный спутник, обладающий атмосферой (состоящей из метана). Остаётся добавить, что принято считать, что шарообразная форма планет и Солнца определяется их гравитационными силовыми полями. Искажение этой формы некоторых планет зависит от их осевого вращения.

Вот, собственно, и все те сведения, которыми мы располагаем о строении планет и Солнца. Они охватывают в общей форме все общепризнанные положения, которыми располагает человечество на сегодняшний день в области внутреннего строения планет и Солнца. У нас есть желание знать больше, чем это известно человечеству. По этой причине мы и отправились в наше путешествие. Увидеть и распознать всё то новое, что встретится в нашем путешествии, поможет нам механика безынертной массы. Теперь остаётся совместить законы и положения механики безынертной массы с условиями состояния планет и Солнца. После чего перед нами откроются картины нового, ранее неизвестного в строении Солнца и планет солнечной системы.

* * *

Начнём с характеристик жидкостей и газов, коль все планеты, в том числе и Солнце, состоят из жидкостей и газов, которые существуют в объёме планеты и её атмосферы в определённом составе. Каждый такой состав отличается от остальных своими химическими, физическими и количественными характеристиками.

Химические характеристики определяют качественную сторону веществ составляющих планеты. Например, что данный газ – азот, данная жидкость есть вода или какое-либо другое соединение. Поскольку мы не собираемся делать химический анализ состава вещества планет, то химическими названиями для жидкостей и газов мы будем пользоваться лишь в отдельных случаях для более чёткого их разграничения.

Из физических характеристик нас будет интересовать, прежде всего, плотность $\rho$ жидкостей и газов. Ибо плотность жидкостей и газов является одной из основных характеристик в зависимостях механики безынертной массы.

Следующей очень важной для нас физической характеристикой будет минимальная энергетическая характеристика веществ, которая определяется давлением $P$ и температурой $T$ этих веществ. Эта характеристика очень старая, но в нашей работе она звучит по-новому. Поэтому прошу уделить ей особое внимание.

Как вы знаете, практически любое вещество может находиться в твёрдом, жидком и газообразном состоянии, так как состояние веществ зависит от температуры и давления среды, в которой находятся эти вещества. Нас же интересует то максимальное давление и та минимальная температура среды, при которых каждое конкретное вещество ещё остаётся в газообразном состоянии. Это значит, что при увеличении температур выше минимальных и при уменьшении давления ниже максимального вещество будет находиться в газообразном состоянии, а при уменьшении температуры ниже минимальной и увеличении давления выше максимального для каждого конкретного вещества оно перейдет в жидкое состояние. Этот энергетический минимум веществ существует как точка, дополненная определёнными вариациями сочетаний давлений и температур. Минимальная энергетическая характеристика вещества нужна нам, прежде всего, для того, чтобы навести границу между жидкостью и газом. Эта характеристика понадобится нам и для других целей, о которых вы узнаете ниже.

### [***Статическая модель планеты***]

Теперь перейдем непосредственно к моделям планет, чтобы совместить положения механики безынертной массы с конкретными условиями существования планет.

Составим модель планеты, которая находится только под действием сил гравитационного силового поля. Эта модель показана на рис. 15 в виде сектора площади сечения планеты [2]. Согласно рис. 15 весь объём планеты разделен на гидросферу и атмосферу. Гидросферу планеты составляют жидкости, которые представляют собой либо определённые элементы таблицы Менделеева, либо соединения этих элементов. В жидком состоянии они находятся лишь потому, что температура среды, в которой они находятся, соответствует их жидкому состоянию.

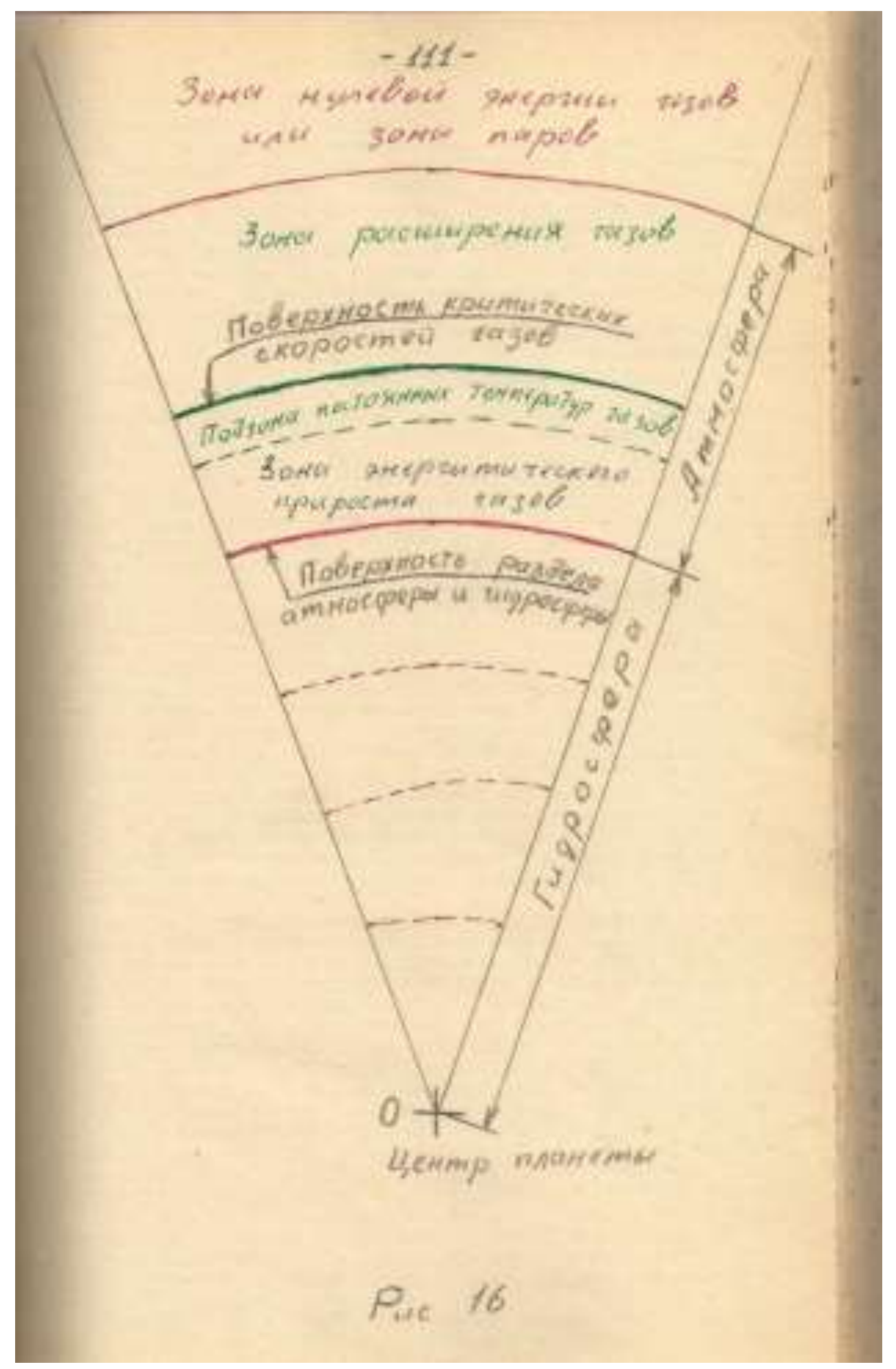

Рис. 15

Атмосфера планеты состоит целиком из газов. Газовый состав атмосферы может быть тоже самым разнообразным. Он определяется соответствующими условиями планеты.

С точки зрения механики безынертной массы вся масса жидкостей и газов планеты находится в состоянии покоя относительно её центра под воздействием гравитационного поля этой планеты. Состояние покоя жидкостей и газов мы определили как состояние застывшего движения. Для планет состояние покоя их массы жидкостей и газов будет определяться двумя видами застывшего механического движения.

Первым видом механического застывшего движения массы жидкостей и газов планеты будет движение этой массы под воздействием гравитационного поля планеты, которое совершается от периферийной границы планеты к её центру по направлению радиуса сферы планеты. Как вы уже знаете, этому движению соответствует постоянная скорость движения $w$, величина которой зависит от величины напряженности силового поля планеты. Для Земли эта скорость $w$ = 3,132 м/сек. В результате этого движения происходит распределение веществ планеты находящихся в жидком и газообразном состоянии по величине их плотности. Вещества с большей плотностью $\rho$ займут положение ближе к центру планеты, а вещества с меньшей плотностью займут положение ближе к периферийной границе планеты. Например, такие элементы, как свинец, ртуть, золото и их соединения, которые обладают большей плотностью в сравнении с другими элементами и их соединениями, займут непосредственно сам центр планеты и прилегающие к нему области. Затем над этими элементами расположатся другие элементы типа кремния, углерода и их соединения. Непосредственно на границе гидросферы с атмосферой расположатся элементы типа натрия, калия и их соединения. На рис. 15 в гидросфере планеты границы веществ с различной плотностью условно показаны штриховыми линиями.

[2] В данной редакции редактор помещает рисунок автора. Поскольку расчёт крыльчатки насоса содержит рисунки, то после его перемещения в работу [1] нумерация рисунков изменилась: рисунок 15 в данном тексте – это рисунок 16 в оригинале.

В атмосфере планеты тоже происходит распределение газов по величине их плотности. Непосредственно у границы поверхности раздела атмосферы и гидросферы будут располагаться более тяжёлые газы типа углекислого, метана или аммиака. Выше этих газов будут находиться газы типа кислорода, азота. Непосредственно на границе планеты будут располагаться самые легкие газы типа гелия, водорода. Границы раздела газов атмосферы планеты на рис. 15 не показаны, но, надо полагать, что подобные границы раздела газов по величине плотности существуют и для атмосферы планеты.

Такое распределение жидкостей и газов по величине их плотности обусловлено тем, что силы давления, создаваемые гравитационным полем планеты, для веществ с различной плотностью будут различны. Это следует из уравнения (122), где показано, что величина сил давления находится в прямой зависимости от плотности вещества. По этой причине более тяжёлые вещества или вещества с большей плотностью вытеснят более лёгкие вещества с меньшей плотностью из центра планеты к её периферийной границе. И мы получаем соответствующее распределение по величине плотности жидкостей и газов в объёме планеты.

Вторым видом механического застывшего движения массы жидкостей и газов будет их движение расширения. Ибо жидкости и газы при более высоких температурах стремятся расшириться и занять больший объём. Это движение жидкостей и газов планеты будет происходить от центра планеты к её периферийной границе по направлению радиуса её сферы, то есть в противоположном направлении к направлению первого вида движения массы вещества планеты.

Чтобы расписать и это движение массы планеты, нам придётся немного коснуться термодинамики. Полагаем, что каждая планета обладает определённым количеством тепловой энергии. Если бы это было не так, то планеты либо остывали, либо, наоборот, разогревались в зависимости от утечек или притока тепловой энергии к планетам. Для звезд тоже будем считать, что количество тепловой энергии для их массы есть величина постоянная. Ибо в этом случае приток и отток тепловой энергии для звезды тоже сохраняется в равных величинах. Тем самым поддерживается постоянство величины тепловой энергии её массы.

Если вы теперь представите себе, что каким-то путем нам удалось изъять всю тепловую энергию планеты, то в этом случае объём планеты уменьшится, особенно объём её атмосферы. Ибо газы перейдут в твёрдое состояние, объём которых будет в тысячи раз меньше, чем объём газообразной атмосферы. Гидросфера уменьшится сравнительно на небольшую величину, так как при переходе из жидкого состояния в твёрдое вещества незначительно уменьшаются в объёме. Следовательно, объём планеты уменьшится до минимально возможных размеров. Если мы снова вернем тепловую энергию планете в полном количестве, то её объём снова займет первоначальное положение. Произойдёт расширение объёма, которое выражается в механическом движении. Поскольку мы рассматриваем состояние покоя массы планеты, то оно как застывшее движение механического движения расширения должно проявлять себя в распределении тепловой энергии в объёме планеты. Тепловую энергию мы определяем по величине температур. Поэтому застывшее движение должно выразиться в неравномерном распределении температур в объёме планеты.

Коль застывшее движение расширения направлено от центра планеты по радиусу к её периферийной границе, то и температуры массы планеты тоже будут убывать от центра к её периферийной границе, то есть в центре планеты её температуры будут максимальны, а на внешней, граничной её поверхности они будут минимальны. В гидросфере планеты это температурное неравенство обеспечивается тем, что в центре планеты располагаются вещества, которые для поддержания своего жидкого состояния требуют максимально высоких температур. Затем над этими веществами расположатся другие вещества, которые требуют для поддержания своего жидкого состояния меньших температур, чем максимальные температуры веществ расположенных в центре планеты. В свою очередь, над этими веществами расположатся  другие вещества с ещё меньшими температурами, и т.д. Наконец, у поверхности раздела атмосферы и гидросферы расположатся вещества, которые требуют минимальных температур для поддержания своего жидкого состояния в сравнении с другими веществами объёма гидросферы.

Отметим, что убывание температур в гидросфере мы указали правильно в общем случае. Это положение легко проверяется на практике. Распределение же веществ в объёме гидросферы по температурам поддержания жидкого состояния мы дали сравнительно приближенно. По той причине, что основным распределением веществ в объёме гидросферы является их распределение по плотности, а плотность веществ и температура их жидкого состояния могут быть различными для каждого вещества относительно его расположения в объёме гидросферы по гравитационному застывшему движению и по застывшему движению расширения. Просто мы здесь отметили, что законами и зависимостями механики безынертной массы мы не можем уточнить или выяснить конкретное распределение температур в гидросфере. Но оно существует как факт, обусловленный тем, что различные вещества имеют различную температуру плавления и в жидком расплавленном состоянии сохраняют эту температуру неизменной. Для атмосферы планеты мы можем дать точные зависимости, которые дадут нам возможность зафиксировать застывшее расширение атмосферы.

Согласно нашей модели, со стороны гидросферы непосредственно по поверхности раздела атмосферы и гидросферы должно располагаться жидкое расплавленное вещество с температурой не ниже 300 – 500°С. Тогда в зоне энергетического прироста газов атмосферы над поверхностью раздела атмосферы и гидросферы расположатся более тяжёлые газы типа углекислого, метана или аммиака. В этом случае в нижней граничной поверхности зоны, которая располагается непосредственно у поверхности раздела атмосферы с гидросферой, температура газа будет равна температуре жидкого расплавленного вещества, которое находится у поверхности раздела атмосферы с гидросферой, то есть 300 – 500°С.

На верхней границе слоя тяжёлых газов температура будет определяться минимальной энергетической характеристикой веществ, которая зависит от давления и температуры. Минимальная же энергетическая характеристика веществ является границей раздела газообразного и жидкого состояния каждого конкретного вещества. В нашем случае такими веществами являются тяжёлые газы. Это значит, минимум температур и давлений верхней границы слоя тяжёлых газов будет определяться этим условием. Ибо если мы снизим температуру границы ниже этого минимума, то тяжёлый газ превратится в жидкость на верхней границе, или если увеличить давление выше этого минимума, то газ тоже обратится в жидкость. Таким образом, мы определили температуры и давления на нижней и верхней границе слоя тяжёлых газов атмосферы.

Зависимость между этими граничными условиями слоя тяжёлых газов или температура и давление в любой точке этого слоя, которая может располагаться на любой высоте между границами слоя, определяется непосредственно самим застывшим движением расширения газов. Ибо застывшее движение расширения газов характеризуется распределением количества тепловой энергии в объёме планеты, в том числе и в атмосфере. По этой причине на долю всего объёма слоя тяжёлых газов атмосферы тоже будет приходиться определённая доля тепловой энергии планеты от общего количества тепла, которое содержится в ней. Это значит, что на объём тяжёлых газов атмосферы приходится определённая и постоянная величина тепловой энергии или постоянное количество тепла. Постоянное и неизменное количество тепла является одним из необходимых условий, из которых мы должны определить промежуточные характеристики тяжёлого газа для его температуры и давления. Для этого нам придётся воспользоваться зависимостями термодинамики. Ибо в этой науке имеются такие зависимости, которые определяют температуру и давление газов в зависимости от механического воздействия. Расширение газов при постоянном количестве тепла ($Q$ = const) в термодинамике определяется зависимостями адиабатического процесса.

Теперь мы получили все необходимые зависимости для физических характеристик газов атмосферы в процессе их расширения. Осталось получить необходимые зависимости для геометрической характеристики застывшего процесса расширения газов. Для этого нам придётся воспользоваться зависимостями механики безынертной массы. Ибо зависимости этой механики отражают механическое воздействие на слой тяжёлых газов. Необходимым условием для этого механического воздействия является первый вид механического застывшего движения газов, или движение газов под воздействием гравитационного поля планеты. Это условие даёт нам возможность, соответственно воспользовавшись, применить зависимости механики безынертной массы (120) - (128) для определения геометрических характеристик застывшего движения расширения газов. Зависимости механики безынертной массы будут последними зависимостями, которые нам необходимы для полной характеристики слоя тяжёлых газов в зоне энергетического прироста газов. Теперь мы получили полный комплекс условий и зависимостей для первого слоя зоны энергетического прироста газов.

Над первым слоем может располагаться один и более слоев других газов по степени непрерывного уменьшения плотности каждого последующего слоя газа относительно предыдущего. Каждый последующий слой газов в этом случае будет определяться теми же условиями, что и первый слой. По этой причине при исследовании каждого из этих слоев мы должны будем поступить точно так же, как мы поступили с первым слоем зоны энергетического прироста газов.

Над вышеописанными слоями газов должен будет расположиться последний слой газа, который обладает повышенным энергетическим минимумом относительно всех этих предыдущих газовых слоев. Здесь мы должны понимать повышенный энергетический минимум в том плане, что все слои газов, кроме последнего, имеют либо плюсовую, либо небольшую минусовую температуру конденсации. Последний же слой газов с повышенным энергетическим минимумом в этом случае должен будет иметь температуру конденсации порядка (–180°С) и ниже. Такими газами могут быть кислород, азот, водород, гелий и т.д. Ибо секрет последнего слоя газов заключается именно в его повышенном энергетическом уровне. Остановимся на этом секрете.

Выше мы установили, что для гравитационного поля каждой планеты существуют свои постоянные скорости застывшего движения. Например, для земного гравитационного поля такая скорость $w = 3{,}132$ м/сек. Теперь снова обратимся к уравнению энергии Бернулли, которое указывает, что полная энергия потока ($VP_{\text{пол}}$) равна сумме потенциальной ($VP_{\text{ст}}$ ) и кинетической ($1/2VP_{\text{дин}}$ ) энергии потока, то есть

$$VP_{\text{пол}} = VP_{\text{ст}} + \frac{1}{2}VP_{\text{дин}}.$$

Из этого уравнения следует энергетическое ограничение при распределении долей потенциальной и кинетической энергий, то есть доля потенциальной энергии должна быть больше или равной доли кинетической энергии потока:

$$VP_{\text{ст}} \geq \frac{1}{2}VP_{\text{дин}}.$$

Коль у нас атмосфера представляет собой поток застывшего движения, то энергетические ограничения связанные с распределением энергий должны в равной степени распространяться и на него с учётом особенностей застывшего потока. Выше мы получили уравнение энергии (128) для жидкостей и газов в векторном силовом поле. Запишем его ещё раз:

$$E_{\text{в}} = V\rho w^2 h.$$

Уравнение (128) означает, что энергия состояния покоя $E_{\text{в}}$ для площади сечения застывшего потока, расположенной на глубине $h$, равна произведению объёма $V$ , который расположен выше площади этого сечения, плотности жидкости или газа $\rho$, квадрату скорости поля планетного тяготения $w^2$ и глубины потока $h$. Нам остаётся только приспособить это уравнение непосредственно к конкретным условиям планеты таким образом, чтобы оно тоже учитывало энергетические ограничения при распределении энергии, как и уравнение Бернулли.

Сначала, чтобы уравнение (128) выглядело проще, отнесём его к единице объёма. Для чего разделим его на объём потока $V$. Получим:

$$P_{\text{пол}} = \rho w^2 h. \qquad (129)$$

Уравнение (129) выражает полную энергию единицы объёма застывшего потока как произведение плотности $\rho$ и квадрата постоянной скорости $w^2$ гравитационного поля умноженного на высоту потока $h$.

Главной особенностью уравнения (129) является то, что оно содержит постоянную скорость $w$, величина которой определяется гравитационным полем соответствующей планеты. В состоянии покоя она лишь помогает выявить количественную величину сил давления и энергии, но если дать возможность жидкостям и газам свободно падать в гравитационном поле планеты, то они будут двигаться именно с этой скоростью. В этом случае застывшая скорость превращается в живую скорость.

Чтобы получить предельное распределение между потенциальной и кинетической энергиями потока в соответствии с уравнением Бернулли, выше мы для этих целей сужали сечение потока до тех пор, пока не получили предельно минимальное сечение потока, в котором реализовывалось это распределение энергий. В данном случае мы тоже имеем дело как бы с суживающимся потоком. Фактически же мы имеем дело с расширяющимся сечением потока. Ибо чем дальше удалена сфера сечения потока от центра планеты, тем площадь её поверхности будет больше. Именно это и происходит с расширяющейся атмосферой планеты с геометрической точки зрения.

Практически же сужение потока выражается в том, что плотность при удалении сечения застывшего потока расширения непрерывно уменьшается, то есть фактическое сужение потока происходит за счёт физического изменения состояния газов. По этой причине на какой-то определённой высоте, когда плотность газов уменьшится до определённого минимума, наша постоянная скорость $w$ станет критической скоростью, при которой происходит предельное распределение энергии в потоке. Для газов эта критическая скорость одновременно является скоростью звука. Скорость звука связана в уравнениях физики и с давлением, и с плотностью. Используя эти зависимости, мы сможем определить и давление, и плотность для критического сечения застывшего потока расширения атмосферы. На рис. 15 эта критическая площадь сечения потока обозначена как поверхность критических скоростей газов.

В этом сечении и статические, и динамические давления потока определяются по уравнению энергий (129). Для живого потока согласно уравнению Бернулли мы должны были бы, чтобы получить полную энергию потока, и потенциальную, и кинетическую вычислить по уравнению (129) и затем сложить. Но поскольку мы имеем дело с застывшим движением, то уравнение (129) будет обозначать сразу три значения: полную, кинетическую и потенциальную энергии потока, которые в нашем случае будут равны друг другу по величине.

Для критического сечения потока кинетическая энергия будет равна как:

$$P_{\text{дин}} = \rho w^2. \qquad (130)$$

Тогда и полная, и потенциальная энергия потока в критическом сечении будут равны той же величине, то есть:

$$P_{\text{пол}} = P_{\text{ст}} = P_{\text{дин}} = \rho w^2. \qquad (131)$$

Уравнения (130) и (131) мы получили из уравнения (129) посчитав высоту $h$ равной нулю. В какой-то степени, зная математику, но ещё не поняв сущности физической картины застывшего движения газов, вы, естественно, возмутитесь и скажете, что умножать на ноль нельзя. Тогда и в результате должен быть ноль, а мы имеем конкретные величины, не равные нулю. Формально вы здесь будете правы, сказав, что высоту $h$ надо приравнять к единице, чтобы получить искомый результат. Но !!!

В работе *Механика жидкости и газа, или механики безынертной массы* мы, при нахождении постоянной скорости движения $w$ жидкостей и газов в поле земного тяготения получили, что высота $h$ является безразмерной величиной. Она просто показывает количество выбранных нами тех или других единиц длины для конкретных условий застывшего движения жидкостей и газов. По этой причине, если бы мы не посчитали высоту $h$ равной нулю, а сделали её равной единице, то мы бы связали единицы измерения длины высоты непосредственно с физическими характеристиками застывшего потока. Мы ведь рассматриваем движение газов непосредственно на поверхности критического сечения потока. Приняв высоту равной единице, мы тем самым задали бы толщину критической поверхности. Поскольку мы не знаем толщину поверхности, то в этом случае мы вынуждены были бы гадать, то ли за единицу измерения взять один километр, то ли один сантиметр. Приняв высоту $h$ равной нулю, мы тем самым приняли, что наша критическая поверхность является началом отсчёта высоты. По этой причине она должна равняться нулю. Тем самым мы не нарушили физической картины движения газов на критической поверхности. Ведь у нас поток всё равно движется со скоростью $w$ при плотности $\rho$. Отсюда следует, что критическое сечение потока, или поверхность критических скоростей газов (см. рис. 15), является началом отсчёта высоты $h$ для всех зависимостей механики безынертной массы, относящихся к состоянию покоя жидкостей и газов в векторном силовом поле. В нашей работе такими зависимостями являются зависимости (120) – (129).

Отметим, что при исследовании физических и им подобных явлений не следует становиться на формальные позиции математических догм, чтобы не дойти до абсурда. Во всех случаях определяющим всегда должна быть непосредственно сама сущность физических явлений, а математику надо рассматривать как средство для оформления этой сущности.

В отличие от живого потока газов, где критическое движение газов реализуется непосредственно в критической площади сечения потока, критическое движение застывшего потока газов реализуется несколько иначе. Ибо для критического сечения застывшего потока расход массы в единицу времени будет тоже застывшей величиной, которая будет выражаться слоем газа соответствующей толщины. В этом слое не происходит адиабатического расширения газов. По этой причине температура во всех точках этого слоя должна быть одинаковой. Энергия этого слоя газа определяется по уравнению (129). Все остальные характеристики определяются соответственно уравнениями (120) – (128). Практически толщину этого слоя можно просто замерить. На рис. 15 толщина критического слоя газов обозначена как подзона постоянной температуры газов. Под нижней границей подзоны постоянных температур газов будет располагаться слой газа того же состава, который находится в подзоне. Толщина этого слоя, как мы выяснили выше, определяется адиабатическим расширением газов в поле земного тяготения. Ещё ниже пойдут слои более тяжёлых газов зоны энергетического прироста газов атмосферы планеты. Для них мы выше уже определили необходимые зависимости.

Далее мы рассмотрим, что происходит с газом выше поверхности критических скоростей газов. В живом потоке, например, в сопле Лаваля, поток газа после критического сечения начинает расширяться. Скорость его непрерывно растет от сечения к сечению, а температура падает, то есть тоже происходит адиабатическое расширение газов. Тепловая энергия преобразуется в кинетическую. Подобное движение газов происходит в застывшем потоке выше критического сечения, но со своими особенностями. Это означает, что выше поверхности критических скоростей происходит непрерывное увеличение застывших скоростей от сечения к сечению за счёт адиабатического расширения газов. В этом случае расширение газов идёт от давления поверхности критических скоростей, которое определяется зависимостью:

$$P_{\text{пол}} = \rho w^2,$$

до нулевого давления. На рис. 15 эта зона обозначена как зона расширения газов.

Фактически же верхняя граница зоны расширения будет определяться минимальным энергетическим уровнем, при котором начинается конденсация газов. Эта минимальная энергетическая характеристика, как мы уже знаем, выражается определёнными минимальными значениями температуры и давления для каждого газа. Расход массы газа в единицу времени для границы конденсации в застывшем движении тоже будет выражаться слоем определённой толщины, который должен быть расположен у верхней границы зоны расширения газов как соответствующая подзона, которая имеет определённую постоянную температуру.

На рис. 15 мы её не показали. Из этих условий мы определили верхнюю границу зоны расширения газов с помощью соответствующих зависимостей. Мы эти зависимости не будем давать, так как они хорошо известны. При необходимости каждый ими может воспользоваться. Коль мы находимся в роли путешественников, то нам сейчас нужна просто общая картина застывшего движения газов атмосферы планеты.

Фактически она будет выглядеть несколько иначе. Ведь застывшее движение расширения газов надо понимать как противодействующее движение механическому движению, которое совершают силы поля планетного тяготения. Не будь этого поля, газы просто-напросто рассеялись бы в мировом пространстве. По этой причине в зоне расширения газов мы имеем не их расширение, а их сжатие полем планетного тяготения, которое тоже происходит в соответствии с адиабатическим процессом. Как вы знаете, при сжатии газов происходит их разогрев. По этой причине газы атмосферы в зоне расширения газов имеют более высокие температуры, чем на поверхности критических скоростей газов. Непосредственно на верхней границе зоны расширения газов газы имеют максимальное значение температур. И по мере приближения к поверхности критических скоростей значение температур непрерывно падает и в своём пределе совпадает с величиной температур поверхности критических скоростей газов. Это будет действительная картина застывшего движения в зоне расширения газов. Ведь выше этой зоны идёт зона нулевой энергии газов, или зона паров. В этой зоне газы полностью теряют тепловую и механическую энергию. На эту зону законы механики безынертной массы не распространяются. Тут будут действовать иные законы физики связанные с электрическими и им подобными свойствами материи.

В зоне нулевой энергии газов газы приобретают способность к ионизации. Поэтому переход из зоны нулевой энергии газов в зону расширения газов будет характеризоваться для газов приростом, или приобретением, механической и тепловой энергии. Поэтому будем считать, что верхняя граница зоны расширения газов одновременно будет границей действия законов и зависимостей механики безынертной массы. Зону нулевой энергии газов или зону паров составляют ионизированные газы. Верхний предел распределения этих газов очень велик. По этой причине мы его искать не будем. Главное в этой зоне то, что она не оказывает никакого силового воздействия на нижние зоны атмосферы.

Добавим, что мы дали построение атмосферы планеты с учётом того обстоятельства, что верхний слой её газов образует застывшее критическое течение с зоной расширения газов с подзоной постоянных температур газов и подслоем адиабатического расширения этих газов. Это происходит в том случае, когда постоянная скорость $w$ поля планетарного тяготения и соответствующее ему давление, плотность и температура газа не превышает энергетического минимума для этих газов. На других же планетах может быть, что состав атмосферы образуют более тяжёлые газы, да и постоянные скорости планетного тяготения тоже могут быть больше, так как они зависят только от поля планетного тяготения. Тогда зависящие от этих скоростей давление, температура и плотность газа могут превысить энергетический минимум для верхнего слоя газов атмосферы этой планеты. В этом случае верхний слой атмосферы не образует критического течения, а ограничивается поверхностью конденсации газов. Тогда строение атмосферы планеты, показанной на рис. 15, претерпит следующие изменения:

поверхность критических скоростей газов превратится в поверхность конденсации газов и станет началом отсчёта высоты $h$ для зависимостей механики безынертной массы;

зона расширения газов исчезнет, а зона нулевой энергии газов будет совпадать своей нижней границей с поверхностью конденсации газов;

подзона постоянной температуры газов будет характеризовать застывший расход массы в единицу времени, который поступает на поверхность конденсации газов.

В остальном зона энергетического прироста атмосферы ничем не отличается от той же зоны, описание которой мы дали выше. Отсюда следует, что верхний слой атмосферы планет может образовывать два вида застывшего движения: критический и конденсационный. Реализация того или другого вида застывшего движения газов верхнего слоя атмосферы для каждой конкретной планеты будет зависеть от состава газа верхнего слоя и от величины сил поля планетного тяготения.

Будем считать, что мы получили необходимое описание модели планеты, масса которой находится в состоянии покоя под воздействием векторного силового поля или поля тяготения планеты. Чтобы окончательно уяснить себе построение модели планеты, повторим это описание в краткой итоговой форме.

Наша модель планеты представляет собой объём шаровой формы. Весь её объём делится на две основные части: гидросферу и атмосферу. Планета обладает определённым количеством тепла, или определённым количеством тепловой энергии. Всё тепло планеты распределяется в её объёме в соответствии с застывшим движением расширения массы планеты. Основное количество тепла планеты содержится в её гидросфере, которая состоит из расплавленных минералов и элементов. Незначительная часть тепла планеты содержится в её атмосфере. В соответствии с застывшим движением расширения массы планеты максимальная температура и максимальное количество тепла находятся в центре планеты. К периферии гидросферы температуры и количество тепла убывают. На граничной поверхности гидросферы и атмосферы планеты гидросфера имеет минимальные температуры и минимальное количество тепла. В атмосфере планетное тепло сохраняется за счёт газообразного состояния самой атмосферы. Распределение температур в газообразных слоях идёт за счёт адиабатического процесса расширения. По этой причине нижние слои атмосферы имеют более высокие температуры, чем верхние. Каждый газовый слой атмосферы ограничивается своим энергетическим минимумом, который определяется температурой и давлением. Верхний слой газа атмосферы тоже ограничивается энергетическим минимумом. В зависимости от величины сил поля планетного тяготения. В верхнем слое атмосферы может реализовываться критическое застывшее движение газов, зона расширения которого тоже ограничивается энергетическим минимумом газов (см. рис. 15). В остальных случаях верхний слой атмосферы ограничивается просто энергетическим минимумом газов. Над атмосферой располагается зона паров, или зона нулевой энергии газов, которая существует в силу иных законов природы.

Если убрать из планеты всё её количество тепла, то вся её масса перейдет в твёрдое состояние и займет соответствующий минимальный объём. Если планету лишить поля её планетарного тяготения, то вся масса планеты перейдет в парообразное состояние и займет максимально возможный объём от воздействия тепловой энергии планеты. Именно в парообразное состояние, так как для каждого газа существуют свои определённые минимумы по давлению в зависимости от его температуры, а пары образуются как ионизированные газы, которые не воспринимают силового воздействия поля планетного тяготения. Это положение мы можем проверить на примере ионосферы нашей планеты.

Тепловому расширению планеты препятствует механическое движение, которое образуется от воздействия поля тяготения планеты направленно - от её периферийных границ к центру. На периферийной границе мы имеем минимальные значения механического давления и энергии, а в центре планеты мы имеем максимальные значения давлений и энергии механического застывшего движения. Это значит, что состояние покоя массы планеты обеспечивается двумя противоположно направленными движениями её массы – это тепловое расширение и противоположное ему механическое движение от действия сил планетного поля тяготения. Механическое движение массы планеты определяется законами и зависимостями механики безынертной массы, а тепловое расширение – законами и зависимостями термодинамики. В этом заключается тесная связь механики безынертной массы с термодинамикой.

**[*Динамическая модель планеты*]**

Далее нам известно, что все планеты солнечной системы имеют осевое вращение, которое мы называем суточным вращением планет. Масса их находится в жидком и газообразном состоянии. По этой причине осевое вращение планет нельзя объяснить законами механики Ньютона. Это значит, что планеты при осевом вращении не подчиняются закону инерции Ньютона, так как жидкие и газообразные тела не сохраняют приданное им вращательное движение. Ибо жидкие тела для своего вращения требуют непрерывной затраты энергии. Это следует из первого закона механики безынертной массы. Это значит, чтобы сохранить осевое вращение жидкого тела, нам необходимо будет подводить к нему энергию весь период сохранения вращательного движения этого тела. Коль планеты состоят из жидкостей и газов, то для их осевого вращения тоже требуется непрерывный подвод энергии. Энергия для жидкостей и газов подводится с помощью соответствующего силового поля. Из четырех видов движения жидкостей и газов вращательным является плоский установившийся вид движения. Для конкретных планетных условий этот вид движения жидкостей и газов будет реализовываться как движение жидкости в центробежном насосе. Ибо силовое воздействие на жидкость планеты при её осевом вращении может быть организовано только в тангенциальном направлении. Тангенциальное силовое воздействие присуще только центробежным насосам при плоском установившемся виде движения жидкостей и газов.

Непосредственно само тангенциальное движение жидкостей планеты не может быть реализовано при помощи колеса с лопатками. В планетах подобное движение жидкостей может быть организовано с помощью соответствующего силового поля. Вы можете подумать, что таким полем может быть, например, магнитное поле Земли. Ибо ионизированные молекулы могут совершать движение под действием магнитного поля. Пожалуй, эта ваша выдумка будет не верна, так как Земля имеет магнитное поле, а Марс его не имеет, но обе эти планеты вращаются почти с одинаковым периодом вращения. Скорее всего, силовое поле планеты, создающее тангенциальное осевое вращение, будет новым, ранее неизвестным человечеству. Поэтому мы не будем говорить, что это за такое силовое поле, а просто отметим как замеченное новое в нашем путешествии. Далее полагаем, что это силовое поле вместо колеса насоса создаёт вращательное движение жидкости в гидросфере планеты.

На рис. 15 мы дали модель планеты, масса которой находится в состоянии покоя под действием двух застывших движений – это движение теплового расширения и движение гравитационного сжатия планеты. Теперь дополним эту модель планеты силовым полем, которое бы создавало плоское установившееся движение в её гидросфере. Естественно, что непосредственно само силовое поле мы не сможем построить и расположить в гидросфере планеты, а вот сам плоский установившийся поток жидкости мы можем получить, исходя из условий и зависимостей для расчёта центробежного насоса[3]. Покажем дополненную плоским установившимся потоком модель планеты на рис. 16.

На рис. 16 показаны два разреза планеты. Верхний разрез сделан плоскостью, в которой размещена ось вращения планеты *O - O*. Нижний разрез сделан по экватору, перпендикулярно оси вращения планеты. На этих разрезах планеты показаны атмосфера, гидросфера планеты и плоский установившийся поток гидросферы. На данном рисунке граница атмосферы планеты определяется либо как верхняя граница зоны расширения газов, либо как поверхность конденсации верхнего слоя газа атмосферы, то есть она определяется как начало отсчёта высоты *h* для зависимостей механики безынертной массы.

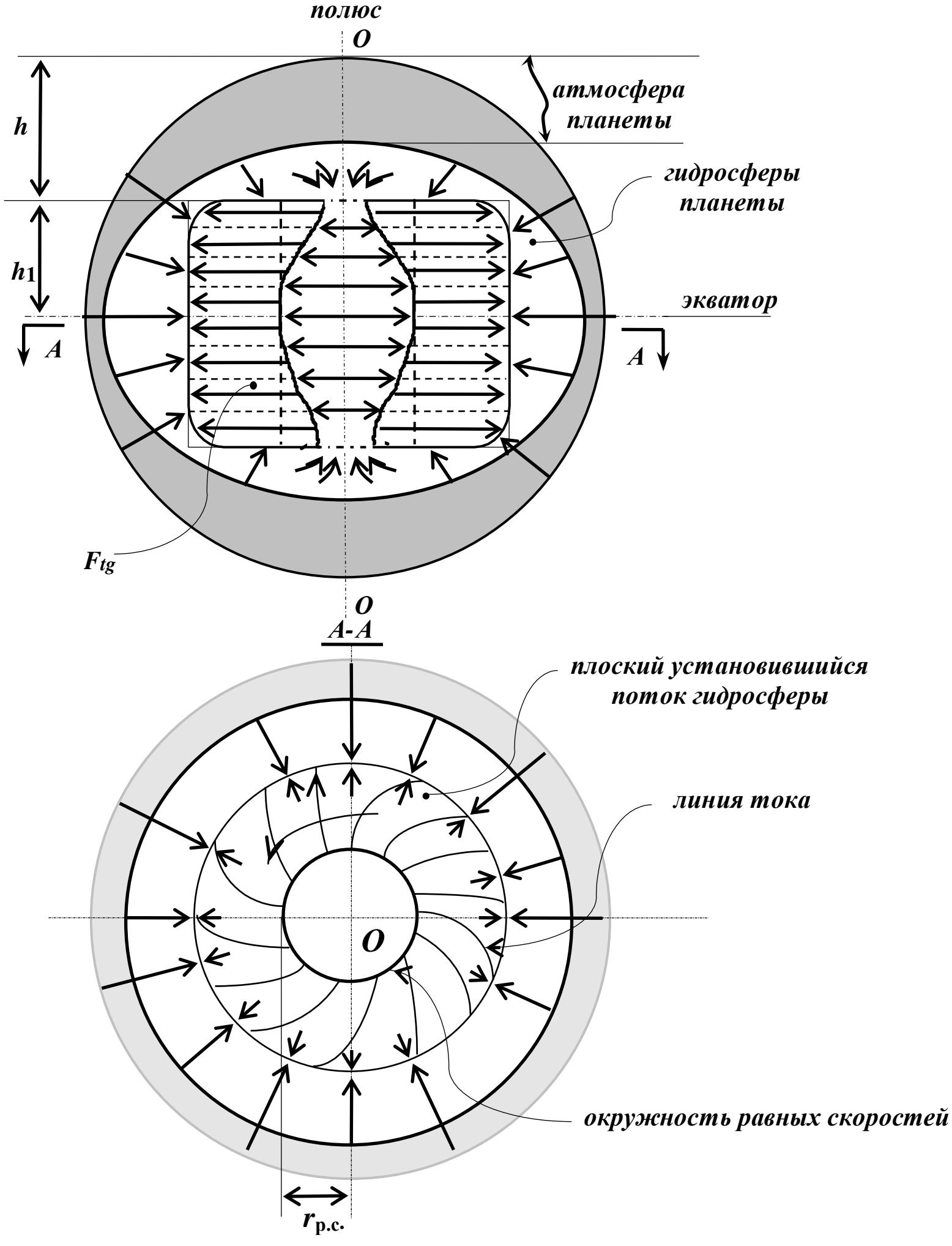

Рис. 16

Построение планеты без плоского установившегося потока гидросферы нам уже известно. Поэтому модель планеты на рис. 16 мы начнём рассматривать непосредственно с плоского установившегося потока гидросферы планеты. Выше мы выяснили, что плоский установившийся поток имеет цилиндрическую форму. На верхнем разрезе (рис. 16) цилиндрическая форма плоского установившегося потока показана штриховыми линиями. В данном случае мы имеем дело, во-первых, с конкретными планетными условиями, которые в конечном итоге определяют форму потока.

Во-вторых, плоский установившийся поток планеты является частным потоком по отношению к общему плоскому установившемуся виду движения жидкостей. Частный характер потока здесь определяется направлением приложения сил в потоке. Направление это будет тангенциальным, как показано стрелкой на нижнем разрезе (рис. 16). Поэтому плоский установившийся поток гидросферы будет потоком насосного типа, или потоком приращения энергии в потоке. Если бы действие сил

[3] См. «Приложение» к работе «Механика жидкости и газа, или механика безынертной массы»

было направлено в радиальном направлении, то мы имели бы дело с другим типом плоского установившегося потока, который относится к турбинному типу, или потоку с убывающей энергией. Поэтому мы будем руководствоваться при построении плоского установившегося потока гидросферы планеты непосредственно конкретными условиями планеты и теми положениями и зависимостями механики безынертной массы, которые относятся к центробежному насосу.

Теперь перейдём непосредственно к плоскому установившемуся потоку жидкости гидросферы. Тангенциальная площадь $F_{tg}$ потока определяется мощностью источника энергии, который создаёт этот поток. В нашем случае источником энергии потока насосного типа является силовое поле. Характеристики этого поля нам не известны, но мы знаем особенности потока, которые должны придаваться потоку этим полем. Минимальный диаметр внутренней граничной поверхности потока будет определяться диаметром окружности равных скоростей. Ибо в данном случае у нас границы потока не определяются какими-то жесткими материальными лопатками колеса насоса, а устанавливаются от действия сил в объёме жидкости. По этой причине ограничивающей внутренней поверхностью потока может быть только поверхность, образованная окружностями равных скоростей. Будем считать, что мы определились с внутренней граничной поверхностью потока в первом приближении.

От этой поверхности к наружной граничной поверхности потока жидкость будет двигаться по поверхностям тока, которые образуют линии тока. Линии тока здесь тоже будут определяться зависимостью логарифмической спирали. Они показаны на рис. 16. Наружную граничную поверхность потока мы бы могли легко определить, если бы знали площадь сечения потока (см. рис. 16).

Вы скажите, что всё здесь начинается с «если бы, да кабы». Раньше ничего подобного не было. Формально вы здесь будете правы, но не больше, чем формально. В данном случае мы тоже решаем задачу, но с той лишь разницей, что условия на проектирование нам не заданы. Для центробежного насоса эти условия были заданы, и мы сравнительно легко решили эту задачу, используя законы и зависимости механики безынертной массы. Коль в нашем случае никаких условий не задано, то нам приходится располагать последовательность решения нашей задачи относительно законов механики безынертной массы. Ибо эти законы утверждают, что причиной осевого вращения планет является плоский установившийся вид движения жидкостей и газов. Далее мы уже делаем как бы обратное решение задачи, так как по существующему потоку стараемся определить его характеристики. Поэтому наши «если бы мы знали то-то, то определили это» являются звеньями взаимосвязи характеристик потока. Они дают возможность восстановить качественную картину движения плоского установившегося потока гидросферы планеты относительно законов механики безынертной массы.

Общая картина движения потока получается не отвлеченная, а вполне конкретная ещё и потому, что мы делаем картину движения потока на основе движения планет солнечной системы, внешняя сторона движения которых нам в определённой степени известна. Поэтому, получив решение задачи в виде взаимосвязи характеристик и условий движения потока, мы получим тем самым качественное решение задачи, которое занимает одинаковую ступень с количественным решением. Ибо подобное решение даёт возможность сопоставить движение планет солнечной системы. К чему мы и стремимся в нашей работе. Также данное решение даёт возможность при минимальных целенаправленных практических исследованиях получить количественные замеры, с помощью которых удаётся получить и количественную картину движения потока гидросферы планеты. Обычно принято считать, что только количественное решение раскрывает тайны того или иного явления природы. Но это не совсем верно. В нашем случае и в других подобных случаях, когда выявлены законы природы определяющие исследуемое явление природы, то качественное решение этого явления, полученное на основании этих законов, даёт возможность как бы увидеть явление природы, а последующее количественное решение даёт возможность как бы пощупать это явление. По этой причине и качественное, и количественное решение существует совместно при исследовании любого явления природы, как глаза и руки. Отсюда следует, что качественным решением мы можем пользоваться как наглядной формой выражения сущности явления природы. Для нашего путешествия этого вполне достаточно. Тем более что сферой нашего путешествия является всего лишь учебник астрономии для школьников.

Выше мы получили геометрические характеристики плоского установившегося потока гидросферы планеты в первом приближении. Нам известно, что поток насосного типа имеет на входе поток жидкости с определённой энергией. Затем к энергии этого потока добавляется энергия работы колеса насоса. Уже на выходе из насоса поток будет иметь энергию равную сумме энергий входного потока и плоского установившегося потока насоса. В нашем случае (см. рис. 16) входной поток жидкости будет располагаться в объёме поверхности образованной окружностью равных скоростей. Далее жидкость входного потока через поверхность равных скоростей поступает в объём плоского установившегося потока. Через внешнюю границу плоского установившегося потока она поступает на выход. Объём выходного потока будет составлять оставшуюся часть гидросферы, расположенную выше внешней границы плоского установившегося потока.

Энергия этого выходного потока должна быть равна сумме энергий входного потока и потока насосного типа гидросферы. Смысл этого энергетического равенства заключается в том, что силы давления объёма выходного потока планеты на внешней границе плоского установившегося потока гидросферы должны будут уравновешиваться силами давления этого потока, которые, в свою очередь, как и энергия, будут состоять из суммы сил давления входного потока и насосного потока. Ибо только в этом случае геометрические границы и характеристики потоков планеты останутся неизменными. Таково условие равновесия.

Мы получили характеристики потоков планеты только с точки зрения центробежных насосов. Теперь нам остаётся уточнить эти характеристики условиями планеты. Планеты имеют шарообразную или близкую к шарообразной форму своего объёма. Объём плоского установившегося потока размещается в этом объёме от плоскости экватора к полюсам планеты (см. рис. 16). Радиусы граничных поверхностей потока располагаются перпендикулярно оси потока планеты. Радиусы поверхностей планеты относительно её оси будут уменьшаться по мере их удаления от плоскости экватора к полюсам планеты из-за шарообразности поверхности планеты. По этой причине радиусы наружной поверхности плоского установившегося потока тоже будут уменьшаться по мере их удаления от плоскости экватора к полюсам планеты.

При этом на каждой величине радиуса поверхности должно сохраняться условие равновесия потоков планеты. На экваторе будет максимальный радиус, у полюсов – минимальный. Наружная поверхность плоского установившегося потока будет иметь форму такую, как приблизительно показано на рис. 16. Из-за уменьшения радиусов наружной поверхности объём потока тоже будет уменьшаться от плоскости экватора к полюсам планеты. Считаем, что напряженность силового поля во всём объёме потока одинакова. Поэтому уменьшение объёма потока, вернее, тангенциальной площади сечения потока, приведет к уменьшению расхода массы в единицу времени у полюсов планеты, поскольку экваториальная часть тангенциальной площади сечения потока будет больше, чем у полюсов.

Величина поверхности равных скоростей находится в прямой зависимости от расхода массы в единицу времени. По этой причине поверхность равных скоростей будет уменьшаться по мере её удаления от экваториальной плоскости к полюсам планеты, то есть её радиус будет уменьшаться. Тогда поверхность равных скоростей примет вид подобный тому, который показан на верхнем разрезе рис. 16. Максимальный радиус окружностей равных скоростей будет располагаться в плоскости экватора, а минимальный – у полюсов планеты.

Нам остаётся ещё уточнить условия равновесия потоков и характер их движения относительно планетных условий. В объёме плоского установившегося потока жидкость движется в любом его тангенциальном сечении с постоянной тангенциальной скоростью. Эта скорость будет одинакова и на внутренней, и на внешней границе потока. Тогда входной и выходной потоки планеты будут соприкасаться с граничными поверхностями плоского установившегося потока, который имеет одинаковые тангенциальные скорости на этих поверхностях. По этой причине оба эти потока будут раскручиваться со скоростью, равной тангенциальной скорости плоского установившегося потока. Объёмы входного и выходного потоков планеты надо рассматривать как жёсткие объёмы. Это значит, что линейные скорости на граничных поверхностях будут различными, а величина их определяется как для жёстких тел вращения. Поэтому объёмы входного и выходного потоков планеты будут находиться в состоянии покоя относительно тангенциального движения жидкости плоского установившегося потока. Тогда энергетический уровень выходного потока будет полностью определяться силовым полем тяготения планеты. Об энергетическом уровне входного потока мы поговорим особо. Для массы объёма входного потока будут действительны уравнения (120 – 128) механики безынертной массы.

Ранее мы рассматривали состояние покоя массы планеты как сумму двух застывших движений: движения теплового расширения и движения от действия гравитационного поля планеты. В данном случае к этим двум застывшим видам движения добавится третий – живой вид движения в виде плоского установившегося потока. Плоский установившийся поток жидкости предусматривает живое движение жидкости в объёме массы планеты. Этот поток насосного типа для своего сохранения требует постоянного расхода массы в единицу времени в тангенциальном направлении движения, который должен соответствовать по мощности источнику энергии. В нашем случае таким источником является силовое поле плоского установившегося потока. Радиальный расход массы в единицу времени для насоса может изменяться в определённом диапазоне. В примере расчёта центробежного насоса мы не рассматривали этого предельного значения изменения радиального расхода массы. Поэтому нам сейчас придётся рассмотреть это изменение.

Полагаем, что на выход нашего расчётного насоса поступает жидкость с полным расчётным расходом и с величиной полного расчётного давления, то есть с расчётной величиной полной энергии на выходе из насоса. Этот расход жидкости обеспечивается радиальным движением жидкости в колесе насоса. Теперь начнём постепенно прикрывать выход насоса до тех пор, пока полностью не прикроем его. Во всех этих случаях полная энергия потока на выходе из насоса остаётся неизменной, то есть расчётной. Это связано с тем, что величина расхода тангенциального потока во всех этих случаях остаётся неизменной, а прирост энергии потока происходит именно за счёт этого потока. Это следует из самого принципа работы центробежного насоса.

Если мы теперь при полностью закрытом входе насоса, когда расход из него равен нулю, уберём входной поток насоса, то полная энергия на выходе из насоса упадет на величину полной энергии входного потока. Насос же будет вращаться и потреблять то количество энергии, которое он потреблял с входным потоком. В этом случае поток на выходе из насоса будет обладать лишь той энергией, которую ему придаёт колесо насоса.

И последний предельный случай. Опять начнём с первоначального положения, когда выход из насоса открыт настолько, что из насоса идёт расчётный расход жидкости с расчётной полной энергией. Теперь поступим наоборот: откроем выход насоса ещё больше, чтобы через него мог пройти увеличенный расход жидкости. В этом случае насос просто выйдет из рабочего режима, то есть движение жидкости не будет связано с плоским установившимся потоком. Это произойдет потому, что ограничителем расхода жидкости через колесо насоса является поверхность равных скоростей. Она определяет максимальный расход жидкости через колесо насоса независимо от того, что внутренняя поверхность на входе в колесо выполнена с большим радиусом, чем окружность равных скоростей. В конечном же итоге всё это зависит от мощности источника энергии насоса. В общем, насос в этих условиях работать не будет.

Мы получили все три предельных случая работы центробежного насоса. Для планет может реализоваться только первый предельный случай работы насоса, который допускает определённый диапазон по изменению радиального расхода.

В этом случае к внутренней граничной поверхности плоского установившегося потока планеты приток жидкости происходит со стороны полюсов планеты (см. верхний разрез рис. 16), который образует входной поток для плоского установившегося потока. Рассмотрим особенности входного потока, исходя из планетных условий.

Полная энергия этого потока должна быть меньше полной энергии выходного потока на величину полной энергии плоского установившегося потока насосного типа. В объёме планеты это условие может быть обеспечено следующим образом. К полюсным отверстиям плоского установившегося потока планеты жидкость будет подходить с энергией эквивалентной высоте $h$ (см. верхний разрез рис. 16). Мы здесь принимаем энергию потока эквивалентной высоте $h$ лишь для наших пояснений. Ведь с ростом высоты увеличивается не только энергия потока, но растет его плотность. Поэтому под эквивалентностью механической энергии высоте $h$ надо понимать комплексное изменение характеристик потока, которое удобнее всего выражать через высоту. Этим удобством мы и воспользуемся для своих пояснений.

В любом случае жидкость к полюсным отверстиям будет подходить с большей механической энергией, чем энергия входного потока, расположенного в объёме поверхности равных скоростей. В результате разницы энергий этих потоков через полюсные отверстия плоского установившегося потока будет поступать определённое количество жидкости, которое в последующем должно пройти через плоский установившийся поток. Это значит, что расход потока в радиальном направлении определяется поступающим через его полюсные отверстия расходом жидкости. Механическая энергия жидкости у полюсных отверстий плоского установившегося потока, эквивалентная высоте $h$, вполне понятна и не требует особых пояснений.

Механическая энергия входного потока, которую мы определили высотой $h_1$, требует дополнительных пояснений. Мы приняли отсчёт высоты $h_1$ от граничной плоскости полюсных отверстий (см. верхний разрез рис. 16). Поэтому мы будем отсчитывать энергию гравитационных сил входного потока от этих плоскостей по высоте $h_1$. Это означает, что величина механической энергии входного потока у полюсных отверстий практически равна нулю, а в плоскости экватора она достигает максимума. Высота $h_1$ входного потока может быть равна почти радиусу планеты. Тогда динамическое равновесие планеты может нарушиться. Вот с какими противоречиями мы сталкиваемся во входном потоке планеты. Это противоречие устраняется, во-первых, тем, что частично максимум механической энергии в плоскости экватора уменьшается количеством энергии эквивалентной радиусу равных скоростей $r_{р.с}$. Ибо гравитационное поле планеты действует во всех направлениях к центру планеты. Поэтому оно будет действовать и в направлении радиуса равных скоростей перпендикулярно действию энергии эквивалентной высоте $h_1$. Это должно быть понятным. Но в этом случае у полюсных отверстий энергия входного потока останется равной нулю.

Во-вторых, для устранения этого противоречия должен производиться разогрев массы жидкости в объёме входного потока. Только в этом случае произойдет уменьшение плотности жидкости в объёме входного потока, и за счёт её теплового расширения произойдет равномерный прирост механической энергии в объёме этого потока. Этот прирост механической энергии во всех точках объёма входного потока произойдет на одинаковую величину. Только в этом случае мы сможем получить необходимое количество механической энергии у входных полюсных отверстий входного потока. Одновременно уменьшится максимум гравитационной энергии в плоскости экватора входного потока за счёт уменьшения плотности жидкости в его объёме, но всё равно этот максимум останется в большей или меньшей степени. При выполнении этих двух условий может быть организован входной поток плоского установившегося вида движения в объёме гидросферы планеты.

Вы заметили, что нам приходится восстанавливать недостающие звенья, например, связанные с изменением тепловых характеристик потока. Вам покажется, что мы затрагиваем чуждые для нас области науки связанные с тепловыми явлениями. Это не совсем верно. Ибо в механике безынертной массы количество тепла содержащееся в массе потока жидкости и газа имеет определённые механические количественные величины, которые выражаются через давление, механическую энергию зависимостями механики безынертной массы. Так что мы имеем прямую и тесную связь между тепловой и механической энергией. Просто нам приходится действовать так же, как лётчику в слепом полёте, когда ему приходится вести свой самолёт только по приборам. Для нас такими приборами являются законы и зависимости механики безынертной массы. По этой причине наши положения относительно моделей планет не относятся к разряду научных гипотез, а являются чисто научными положениями. Просто они требуют количественного уточнения границ потоков планеты. Сейчас мы этого не сможем сделать, но в будущем мы получим такие количественные уточнения. Их можно, например, получить путем замеров так называемого ядра планеты, которое у нас именуется входным потоком. Получив такие замеры, мы сможем сразу же получить внешние границы потока. В земных условиях мы можем создавать планетные потоки на действующих моделях. В последующем модельные установки могут пригодиться для практических целей. Ибо в наше время ещё плохо изучены сверхвысокие температуры. Как знать, может быть, подобные установки дадут возможность получить энергию термоядерных реакций для практического её использования в энергетическом хозяйстве людей.

Теперь ещё раз остановимся на динамическом равновесии потоков планеты, чтобы свести их в единую картину.

Плоский установившийся поток располагается в гидросфере планеты согласно рис. 16. Источником энергии для этого потока служит соответствующее силовое поле, которое придаёт потоку определённое количество энергии. Это количество энергии придаёт потоку точные геометрические характеристики. Точность геометрических характеристик потока выражается в том, что тангенциальная площадь сечения потока по своей величине остаётся неизменной, или постоянной, для каждого конкретного силового поля, если даже по каким-либо причинам изменится конфигурация этого сечения. Ибо тангенциальная площадь сечения потока соответствует всегда количественному притоку механической энергии.

Внутренняя граничная поверхность плоского установившегося потока образуется окружностями равных скоростей. В плоскости экватора окружность равных скоростей имеет максимальный диаметр, а у полюсов планеты она имеет минимальный диаметр. Если эта разница по диаметрам окружностей равных скоростей не приводит к различию угловых скоростей по высоте плоского установившегося потока, то этот поток будет иметь у полюсов планеты меньшее количество энергии за счёт уменьшения в этих местах количества участвующей в движении массы жидкости потока. То есть различие энергий в потоке определяется конфигурацией тангенциальной площади сечения потока, а не источником энергии. Если разница по диаметрам окружностей равных скоростей приводит к различию угловых скоростей по высоте потока, то это будет связано с различием источников энергии. Это значит, что по высоте потока располагается несколько силовых полей с различными энергетическими уровнями. Например, некоторые планеты у полюсов вращаются медленнее, а в области экватора имеют более быстрое вращение. Это значит, что к полюсам энергетический потенциал силового поля уменьшается. Поэтому плоский установившийся поток у полюсов имеет меньшие тангенциальные скорости, чем в области экватора.

В объёме поверхности равных скоростей располагается входной поток плоского установившегося движения гидросферы планеты. В объёме этого потока происходит разогрев содержащейся в нем массы до сверхвысоких температур. Будем считать, что источником тепловой энергии являются термоядерные реакции. Ибо из всех известных нам источников тепловой энергии только термоядерный источник способен обеспечить разогрев массы входного потока до сверхвысоких температур. В свою очередь тепловая энергия входного потока обеспечивает необходимый уровень его механической энергии, от величины которой зависит расход массы в единицу времени через полюсные отверстия потока. Ибо эта энергия создаёт определённый перепад по давлению с выходным потоком планеты, который создаёт соответствующий расход массы через полюсные отверстия планеты.

Некоторое количественное различие механической энергии в объёме входного потока по его высоте от полюсных отверстий к плоскости экватора создаётся силовым полем тяготения планеты. По этой причине у полюсных отверстий входного потока мы наблюдаем минимум механической энергии, а в плоскости экватора планеты – максимум. Это различие в механической энергии в объёме потока может быть ликвидировано за счёт увеличения радиусов внутренней граничной поверхности в области максимальных энергий, то есть в области экваториальной плоскости. Плоский установившийся поток насосного типа допускает увеличение радиусов внутренней граничной поверхности потока больше радиусов окружностей равных скоростей. Это мы знаем из примера расчёта центробежных насосов. По этой причине в экваториальной зоне входной поток будет иметь граничную поверхность с большими радиусами, чем соответствующая поверхность равных скоростей. Поэтому граничная поверхность потока перейдет в новое положение. В результате чего произойдет выравнивание механической энергии в объёме входного потока с одновременным увеличением его объёма. При этом величина тангенциальной площади сечения плоского установившегося потока должны остаться неизменной.

Если после такого выравнивания механической энергии в объёме входного потока произойдет дополнительное увеличение температуры массы этого объёма, то в этом случае должны будут увеличиться радиусы внутренней граничной поверхности потока по его высоте от полюсов до экватора сверх тех величин, которые мы имели при выравнивании энергий. В результате чего произойдет новое увеличение объёма входного потока. При всех этих увеличениях и изменениях динамическое равновесие всех потоков планеты должно сохраняться. По этой причине, например, круглое сечение планеты может превратиться в элипсное.

Главной особенностью входного потока остаётся то, что в его объёме происходит разогрев массы до сверхвысоких температур. В противном случае невозможно существование плоского установившегося потока в объёме планеты.

Из входного потока жидкость поступает через внутреннюю граничную поверхность в объём плоского установившегося потока. В этом потоке происходит прирост механической энергии до соответствующих величин. В результате сжатия массы в потоке увеличивается её температура. Затем эта масса жидкости в виде радиального расхода через наружную граничную поверхность плоского установившегося потока насосного типа поступает в выходной поток. Энергия выходного потока определяется силовым полем планетного тяготения и зависит от высоты $h$. Масса жидкости выходного потока имеет более низкие температуры, чем жидкость плоского установившегося потока. Уменьшение температур в массе жидкости этого потока от внешней границы плоского установившегося потока до граничной поверхности атмосферы планеты идёт в соответствии с застывшим тепловым расширением. Максимум температур находится на внешней граничной поверхности плоского установившегося потока, а минимум температур – в атмосфере планеты. Осевое вращение выходного потока определяется тангенциальной скоростью плоского установившегося потока планеты. Поэтому относительно тангенциальной скорости выходной поток находится в состоянии покоя. Это значит, что видимое нами суточное вращение планеты определяется именно тангенциальной скоростью установившегося потока гидросферы планеты. Частично жидкость из выходного потока планеты поступает через полюсные отверстия входного потока плоского установившегося потока планеты. Затем эта жидкость через плоский установившийся поток снова возвращается в выходной поток. Таким образом происходит непрерывная циркуляция массы планеты в пределах её объёма. Вот, собственно, и вся картина уточнённого динамического равновесия планеты.

[***Совмещённая модель планеты***]

Мы рассмотрели как бы две модели состояния планет. Одна из них связана с застывшим движением теплового расширения и застывшим механическим движением поля планетарного тяготения. Другая связана с динамическим равновесием потоков планеты. Но существование планет связано непосредственно с совмещённым состоянием, когда эти две модели представляют собой одну общую модель со всеми своими застывшими и динамическими движениями. Теперь мы будем исходить из этой одной общей модели. Для совмещения этих моделей мы воспользуемся законом сохранения энергии, который был получен М. В. Ломоносовым. Это значит, что существование всех планет определено наличием механической энергии поля планетного тяготения, которое уравновешивает энергию застывшего теплового расширения планеты и энергию плоского установившегося потока гидросферы планеты.

Поскольку тепловая энергия в механике безынертной массы воспринимается через механическую энергию, то мы будем вправе, в конечном итоге, рассматривать закон сохранения энергии для планет в механической форме. Тепловая энергия как бы косвенно участвует в механической энергии через изменение плотности массы. Поэтому для планет мы должны будем приложить закон сохранения энергии к плоскому установившемуся потоку и к застывшему движению поля планетного тяготения как к источникам механической энергии.

Энергия плоского установившегося потока стремится разбросать массу планеты, а энергия гравитационного поля планеты стремится воспрепятствовать этому. Энергия гравитационного поля планеты может противодействовать энергии плоского установившегося потока лишь в том случае, когда она либо больше, либо равна энергии этого потока.

Энергия плоского установившегося потока планеты сосредоточена во входном потоке и непосредственно в самом плоском установившемся потоке. Энергия гравитационного поля сосредоточена в объёме планеты, который располагается от внешней граничной поверхности плоского установившегося потока до граничной поверхности атмосферы планеты. Динамическое равновесие планеты определяется количественным равенством между энергией потока $E_{\text{п}}$ и энергией объёма планеты – сосредоточения гравитационной энергии $E_{\text{об.т}}$, то есть

$$E_{\text{п}} = E_{\text{об.т}}. \qquad (132)$$

Только при сохранении этого равенства энергий возможно существование плоского установившегося потока насосного типа в гидросфере планеты, то есть оно является обязательным условием.

Если мы уберём с планеты гравитационную энергию, то планета не будет существовать. Она просто распадется. Если мы уберём из планеты плоский установившийся поток, то она будет продолжать своё существование как самостоятельное космическое тело. Максимальный уровень механической энергии поля планетного тяготения $E_{\text{т}}$ для такой планеты будет расположен в её центре. Количественную величину этого уравнения мы можем подсчитать. Ведь нам известна высота $h$ для каждой планеты, которая в данном случае будет равна радиусу планеты. Средняя плотность $\rho$ массы планеты нам тоже известна. Подставим эти величины в уравнение (128) и мы получим искомую величину максимальной механической энергии поля планетного тяготения. Согласно закону сохранения энергии эта максимальная величина поля планетного тяготения есть величина постоянная для каждой планеты, то есть

$$E_т = \text{const}.$$

Это положение является обязательным условием для любой планеты. Оно означает, что как бы не преобразовывалась механическая энергия внутри объёма планеты, сумма её должна быть всегда постоянной и равной максимальной, или полной, энергии $E_т$ поля планетного тяготения. Согласно этому условию, для планет, в объёме которых размещается плоский установившийся поток, сумма энергий этого потока и остального объёма планеты, в котором сосредоточена энергия поля планетного тяготения $E_{об.т}$, должна быть равна максимальной, или полной, энергии поля планетного тяготения, то есть

$$E_т = E_п + E_{об.т}. \qquad (133)$$

Согласно первому условию (132), энергии плоского установившегося потока и оставшегося объёма планеты равны между собой. Это значит, что каждая из этих энергий должна равняться половине полной энергии планеты.

После этих условий все наши качественные картины моделей планет приобретут количественный смысл. Как мы показали выше, полную механическую энергию планеты мы можем всегда получить, даже пользуясь средней величиной плотности массы планеты, не зависящей от величины давления. Если мы дальше примем для моделей планеты среднюю плотность массы, то мы сможем найти внешнюю границу плоского установившегося потока. По условию равенства энергий она должна находиться на середине радиуса планеты, то есть радиус внешней границы плоского установившегося потока гидросферы в плоскости экватора будет равен половине радиуса планеты. Например, для нашей планеты Земля при тех же условиях этот радиус будет равен 3 000 км. По радиусу мы сможем найти длину окружности. Эта длина окружности будет соответствовать пути точки, которая та должна проделать за 24 часа, то есть за земные сутки. Зная время и путь, мы определим скорость. Эта скорость будет соответствовать тангенциальной скорости плоского установившегося потока гидросферы планеты. При тех же условиях для земного плоского установившегося потока она будет равна порядка $W_{tg}$ = 250 м/сек. Если принять величину средней земной плотности равной 5,5 г/см$^3$, то величина сил давления в этом потоке будет порядка 30000 кг/см$^2$. В общем, в такой последовательности мы сможем определить все характеристики динамического равновесия планеты в количественном виде. Для действительных планет все эти характеристики необходимо будет получать с учётом сжимаемости массы планеты и некоторых других дополнительных условий, которые были изложены в качественных картинах моделей планет. По этой причине радиус внешней границы плоского установившегося потока будет иметь иное количественное значение, чем при средней величине плотности массы планеты. Это иное количественное значение определяется уже не как ошибка, а как степень точности расчётов.

Для нас эта точность расчёта вполне достаточна, чтобы выявить визуальные различия планет. Кому потребуется более высокая степень точности расчётов, тому придётся проделать ряд дополнительных исследований связанных с изучением сжимаемости массы при сверхвысоких давлениях и температурах. Нам же подобных исследований не провести, коль у нас нет ничего, кроме учебника астрономии. На этом можно было бы закончить о моделях планет, но мы ещё не рассмотрели предельные условия для их динамического равновесия.

Мы рассматривали первый случай, когда механическая энергия плоского установившегося потока насосного типа размещённого в объёме гидросферы планеты уравновешивается механической энергией поля планетного тяготения, которую содержит масса объёма планеты, расположенного выше внешней граничной поверхности плоского установившегося потока. Назовём подобное состояние планет состоянием динамического равновесия.

Второй предельный случай бывает тогда, когда механическая энергия поля планетного тяготения превышает механическую энергию плоского установившегося потока планеты. Практически этот случай реализуется в планетах в таком виде:

уровень механической, или полной, энергии поля планетного тяготения размещается в центре планеты. В этом случае энергетический уровень планеты полностью определяется энергией поля планетного тяготения. Плоский установившийся поток размещается в объёме планеты как простое течение, которое не изменяет энергетического уровня планеты. Например, как течение в океане, которое не нарушает энергетический уровень самого океана. Это значит, что состояние покоя такой планеты мы должны будем определить по первой модели как совокупность застывших движений: теплового расширения и механического движения поля планетного тяготения. Здесь мы должны будем пользоваться всеми теми условиями, которые мы получили для первой модели планеты.

При наличии на этой планете силового поля, которое создаёт плоский установившийся поток, в ней тоже организуется этот поток. Но количественным ограничением плоского установившегося потока в этих условиях будут скорости. Это значит, что величина тангенциальной скорости $W_{tg}$ этого потока должна быть меньше величины постоянной скорости поля планетного тяготения $w$, то есть

$$w > W_{tg} .$$

В этом случае плоский установившийся поток будет существовать в объёме планеты как течение, которое не нарушает общего энергетического уровня планеты связанного с силовым полем планетного тяготения.

Внутренняя граничная поверхность этого потока-течения будет тоже определяться поверхностью равных скоростей. Здесь она будет означать внутреннюю границу течения, которая уже не является одновременно границей энергетического различия потоков. Внешняя граница этого потока-течения, коль она не определяет энергетического различия, будет совпадать с наружной границей гидросферы планеты. Для этих планет их осевое вращение тоже будет определяться величиной тангенциальной скорости плоского установившегося потока-течения. Поэтому величину тангенциальной скорости для этих планет мы можем определить по наружному диаметру гидросферы планет в плоскости экватора. Назовем второй предельный случай состояния планет состоянием статического равновесия с динамическим движением.

При третьем предельном случае, когда энергия плоского установившегося потока превышает энергию поля планетарного тяготения, происходит разрушение планеты. Практически же планета не разрушается, а происходит отстрел спутника. Тем самым снижается полная энергия планеты, и она продолжает своё существование с новым энергетическим уровнем и со своим спутником. Это явление совершается тоже по законам механики безынертной массы. Оно поддаётся количественному и качественному решению, но требует дополнительного изучения. В настоящее время мы не располагаем необходимыми возможностями для изучения этого явления. Поэтому отложим его до лучших времен. Назовём это третье состояние планет состоянием изменения энергетического уровня.

После того, как мы получили три предельных случая состояния механического равновесия планет, мы теперь можем получить предельные случаи распределения тепловой энергии в объёме планеты.

Здесь мы будем исходить из тех позиций, что при увеличении давления увеличивается тепловая энергоемкость жидкости. Например, мы знаем, что с увеличением давления растет температура кипения воды. Температуру мы понимаем как показатель теплового энергетического уровня жидкостей и газов. Как манометрическое давление является показателем механической энергии жидкостей и газов, так и температура является подобным показателем для уровня тепловой энергии жидкостей и газов.

Мы представляем себе гидросферу планеты как определённый объём заполненный массой обладающей свойствами жидкости. Всякая жидкость при определённом давлении имеет конкретный максимум температур, который для неё остается постоянной величиной. Это значит, что выше этого максимума мы не сможем поднять температуру этой жидкости, как бы мы её не нагревали. Дальнейший подвод тепла лишь приведёт к испарению жидкости, но не к повышению её температуры. Мы, конечно, не знаем, как ведут себя жидкости при сверхвысоких температурах и давлениях, но мы хорошо знаем, что температура жидкости находится в прямой зависимости от давления. Всё это надо понимать для планетных условий следующим образом:

что источник тепловой энергии планеты может иметь сколь угодно большие температуры. Этот источник отгораживается от остальной части объёма планеты непосредственно прилегающим к нему слоем жидкости с определённым давлением. Поэтому этот слой жидкости вбирает в себя от источника тепловой энергии планеты такое количество тепла, которое может вобрать в себя этот слой жидкости при соответствующем давлении, то есть если температура источника тепловой энергии была бы бесконечно большой величиной, то слой жидкости будет иметь вполне конкретную величину температуры. Величина этой температуры будет тем больше, чем больше давление этого слоя жидкости. Затем этот прилегающий к источнику тепла слой жидкости будет передавать тепловую энергию следующему прилегающему слою жидкости уже относительно своей температуры, а не температуры источника тепла. Температура этого последующего слоя должна быть ниже, так как давление в этом слое будет ниже. Это значит, что при определённой мощности теплового источника, который расположен в центре планеты, температура жидкости гидросферы планеты будет уменьшаться от центра планеты к её периферийной границе по причине уменьшения давления в слоях жидкости гидросферы в той же последовательности.

Для наших планетных условий смысл этой последовательности распределения тепловой энергии заключается в том, что, если мы возьмём две одинаковые по объёму и по массе планеты с одинаковой мощностью источников тепловой энергии расположенных в центрах планет, но одна из этих планет будет находиться в состоянии динамического равновесия, а вторая – в состоянии статического равновесия с динамическим движением или без него и для этих двух планет максимум механической энергии будет одинаков, то различие здесь будет заключаться в том, что в первой планете источник тепловой энергии будет окружен слоями жидкости с меньшим механическим давлением, а во второй планете – с большим механическим давлением. Ибо для первой планеты в центре её будет располагаться минимальное давление плоского установившегося потока гидросферы планеты, а во второй планете в её центре будет располагаться максимум давлений поля планетного тяготения. Это значит, что энергоёмкость или температура в объёме плоского установившегося потока гидросферы первой планеты будет ниже, чем у второй планеты. Практически такое различие по температуре достигает сотен градусов. Например, для нашей планеты, если бы мы её перевели из состояния динамического равновесия в состояние статического равновесия, это различие означало, что наши материки превратились бы в раскаленные глыбы, либо вовсе расплавились.

В общем, все наши рассуждения о тепловом распределении в объёме планеты мы будем сводить к тому, что направление движения плоского установившегося потока в объёме гидросферы планеты совпадает с направлением её застывшего теплового движения. Тепловая энергия, как вы уже знаете, преобразуется в механическую. Если бы тепловая энергия объёма плоского установившегося потока соответствовала тепловой энергии в том же объёме, но для планеты статического равновесия, то суммарная энергия - тепловая и плоского установившегося потока, намного превысила бы полную энергию планеты. Тогда бы мы имели противоречие с законом сохранения энергии. По этой причине в объёме плоского установившегося потока гидросферы температура должна снижаться соответственно. Отсюда следует, что часть тепловой энергии источника тепла планеты преобразуется в кинетическую энергию плоского установившегося потока гидросферы планеты при её состоянии динамического равновесия.

Отметим, что в определённых случаях тепловая энергия в объёме плоского установившегося потока может быть больше нормы. В этом случае излишки механической энергии гасятся за счёт деформаций или дополнительного расширения объёма плоского установившегося потока. По этой причине объём планеты деформируется. Осевое её сечение из круглого превращается в элипсное. Но величина этих деформаций тоже определяется законом сохранения энергии и допустимыми пределами механического равновесия между плоским установившимся потоком гидросферы планеты и её остальным объёмом со статической энергией поля планетного тяготения. Вот так мы будем понимать распределение энергии в объёме планет в зависимости от состояния их механического равновесия. Уменьшение плотности вещества планеты в объёме плоского установившегося потока при её состоянии динамического равновесия будем понимать как преобразование тепловой энергии источника тепла планеты в кинетическую энергию плоского установившегося потока в этом объёме. За счёт чего происходит уменьшение общего температурного режима в объёме планеты соответственно.

В данной работе мы были и остаёмся всего-навсего путешественниками, которые желают увидеть для себя новое в этом путешествии. Так вот, чтобы увидеть это новое, мы должны представлять его себе в таких конкретных формах, которые позволили бы нам наглядно увидеть интересующие нас различия. Эти различия определяются законами механики безынертной массы. Они выражают только общие различия присущие всей материальной природе. Три предельных состояния механического равновесия планет и связанное с ними распределение тепловой энергии в объёме планеты, являются теми концентрированными различиями, которые мы в состоянии удержать в своей памяти и наглядно представить себе. А уж эти состояния планет дают нам возможность делать необходимые наблюдения для выявления нового. Чтобы получить эти наглядные формы для планет солнечной системы, мы изучили с вами законы и зависимости механики безынертной массы, рассчитали центробежный насос, составили модели планет. Другими словами, всё это называется – внести ясность в интересующий вас вопрос. Как видите, сама ясность выражается в немногом, но очень ёмком, а предварительная работа для выявления этого немногого требуется очень большая. Ведь на пути ко всякой ясности лежит столько каверз и каждую из них надо распутать и привести в определённую систему. Способностью получать ясность в исследуемых вопросах определяется ценность любого специалиста. Вот для каких целей вам нужны ваши школьные знания. Для вас данная работа в первую очередь должна быть примером применения этих знаний к известной для вас жизненной практике.

**[*Строение атмосферы и гидросферы планет земного типа*]**

Имея наглядную форму для различения нового, мы теперь можем приступить к разбору планет солнечной системы. Начнём с ближайшей к Солнцу планеты. Этой планетой является Меркурий. Общие характеристики планеты нам известны. Мы их взяли из учебника астрономии. Далее начинаем рассматривать эту планету с точки зрения механики безынертной массы.

Нам известно, что Меркурий имеет медленно осевое вращение. Согласно нашим положениям это означает, что объём массы планеты находится в состоянии статического равновесия с динамическим движением и что в центре планеты размещается тепловой источник. По этой причине температура поверхности планеты имеет несколько сот градусов. Это подтверждается тем, что измеренная на подсолнечной стороне температура – более 300°С. С нашей планеты Земля можно измерить только температуру на поверхности критических скоростей газов атмосферы планеты. Поэтому измерения температуры относятся именно к этой поверхности. В связи с тем, что гравитационное поле Меркурия меньше земного, то зона энергетического прироста газов его атмосферы будет иметь высоту всего несколько километров. Из-за этого разница по величине температур между поверхностью критических скоростей газов атмосферы и поверхностью планеты, или поверхностью раздела атмосферы и гидросферы (см. рис. 15), будет исчисляться как 100–200°С. Поэтому поверхность Меркурия будет иметь температуру порядка 400–500°С, то есть мы для этого должны к измеренной температуре (300°С) прибавить разницу температур. Температуру поверхности и высоту зоны энергетического прироста газа атмосферы можно вычислить точно, если знать состав газа атмосферы. Мы этого состава не знаем, поэтому эти величины подсчитать не сможем.

**Следующей планетой, которую мы осмотрим, будет Венера.** Она интересна нам в первую очередь тем, что на её поверхности уже побывали автоматические станции. Что позволит нам в определённой степени получить экспериментальное подтверждение нашим положениям.

Гидросфера Венеры находится в состоянии статического равновесия с динамическим движением. В её центре располагается источник тепловой энергии. Температура на поверхности Венеры около 500°С, замеренная автоматическими станциями, подтверждает наше положение о том, что при одинаковой мощности тепловых источников планет температура поверхности или тепловая энергоёмкость у планет в состоянии динамического равновесия меньше, чем у планет в состоянии статического равновесия. В состоянии статического равновесия механическая энергия планеты определяется её полем тяготения. Плоский установившийся поток в гидросфере Венеры сохраняется в виде течения, которое не изменяет механического энергетического уровня. Он тоже является следствием определённого силового поля планеты. За счёт этого течения Венера имеет определённое осевое вращение. Как мы знаем, один оборот вокруг своей оси Венера совершает за 243 суток. В связи с тем, что энергетический уровень планеты определяется её гравитационным полем, строение её гидросферы должно быть слоистым. Это значит, что к центру планеты будут располагаться слои её массы с большим объёмным весом, а к поверхности – с меньшим объёмным весом. Поэтому поверхность Венеры находится либо в жидком состоянии, либо она образована твёрдым пористым веществом в виде пемзы, температура плавления которого выше, чем температура поверхности Венеры. Если такая планетная кора существует, то она имеет небольшую толщину, которая измеряется либо десятками, либо сотнями метров. Вот чем мы можем пополнить наши знания о строении гидросферы Венеры.

Как пишут, атмосфера Венеры состоит на 95% из углекислого газа. Остальное её содержание составляет азот, кислород и инертные газы. Строение атмосферы планеты Венера возьмём из журнала «Наука и жизнь», № 4 за 1972 год. В статье этого журнала, которая называется «Вода на планетах», авторов В. Дерпгольц и Г. Каттерфельда, приводится таблица атмосферы Венеры. Покажем её на рис. 17. Согласно рис. 17, начиная с высоты 55 км и до высоты 60 или 65 км располагается слой облаков водяных паров. Полагают, что облака составляют не пары, а ледяные кристаллики микронного размера. Толщина слоя облаков 7 – 10 км. Таблица атмосферы Венеры, как утверждают авторы, построена по данным «Венеры – 4», «5», «6», «7» и «Маринера – 5».

Теперь мы эти данные обработаем со своей точки зрения. Для этого нам придётся вспомнить рис. 15, где показано модельное построение атмосферы планеты. Тогда атмосфера Венеры для нас будет смотреться в таком плане. Непосредственно над поверхностью Венеры будет располагаться зона энергетического прироста газов её атмосферы. Эта зона состоит из двух слоев газов. Первый углекислотный слой газа располагается от поверхности Венеры до высоты 50 км. Над этим слоем располагается второй слой азотно-кислородного состава. Он ограничивается поверхностью критических скоростей газов. Она располагается на высоте 65 – 70 км от поверхности Венеры. Под этой поверхностью располагается подзона постоянной температуры газов. Толщина этой подзоны должна быть в пределах 7 – 10 км. Выше располагается зона расширения газов. Её верхняя граница находится на высоте порядка 100 км. Верхняя граница зоны расширения газов одновременно является границей распространения действия законов механики безынертной массы в атмосфере Венеры. Ибо все зоны атмосферы планеты по нашей классификации характеризуются определёнными формами застывшего движения.

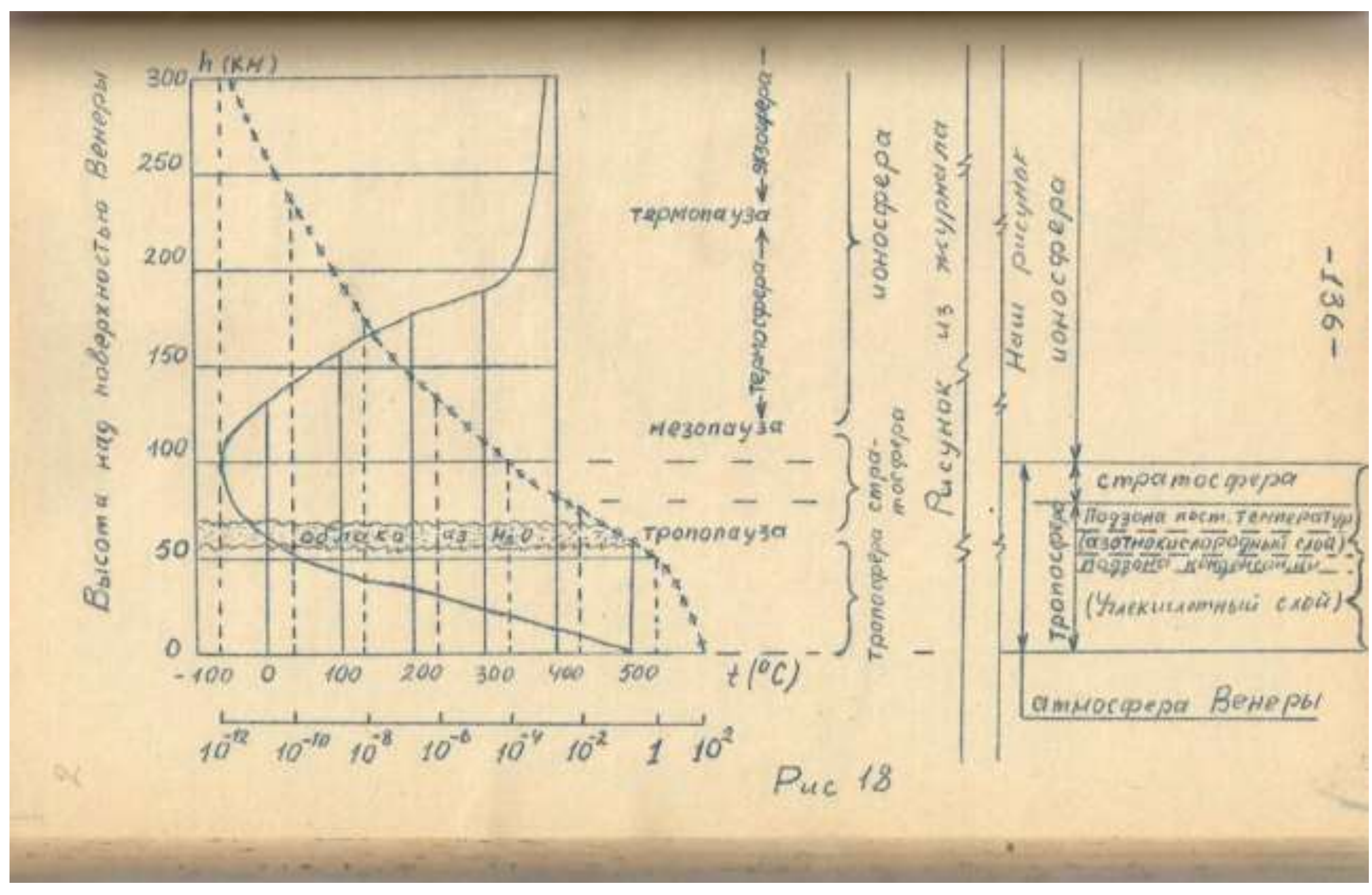

Рис. 17

Так зона энергетического прироста газов отражает застывшее движение адиабатического расширения газов. Подзона постоянных температур газов характеризует застывший критический расход газов через поверхность критических скоростей газов. Одновременно эта поверхность является началом отсчёта постоянных скоростей *w* поля планетного тяготения. Зона расширения газов характеризует собой застывшее движение расширения газов до минимально возможного уровня механической энергии этих газов. Этот минимум механической энергии определяется по конденсации газов. Выше зоны расширения газов на рис. 15 показана зона нулевой энергии газов, или зона паров. Здесь под нулевой энергией имеется в виду механическая энергия, которая отсутствует в этой зоне. По этой причине законы механики безынертной массы не распространяются на неё. Поэтому мы ограничили атмосферу планеты зоной расширения газов.

Здесь нам приходиться пользоваться понятиями газа и пара, но мы до сих пор не определились с этими понятиями, то есть не внесли ясность в этот вопрос. Под газами мы понимаем любое химическое вещество, для которого состояние между тепловой и механической энергиями определяется законами механики безынертной массы.

Под парами мы понимаем состояние любых химических веществ, подобное их газообразному состоянию, которое характеризуется отсутствием механической энергии. Это значит, что пары не подчиняются законам механики безынертной массы, а подчиняются каким-то иным законам в своём механическом движении и состоянии. Например, для воды мы называем паром всё: это пар в паровых котлах электростанций, который находится под давлением от нескольких атмосфер до сотен атмосфер, пар, который поднимается от кипящей кастрюли и с поверхностей рек, озер и т.д. С точки зрения механики безынертной массы водяной пар находящийся под давлением является не паром, а водяным газом. Пар выходящий из кипящей кастрюли и пар поднимающийся с поверхностей озер и рек является паром с точки зрения механики безынертной массы. Это значит, что атмосферное давление Земли для паров веществ испаряющихся с открытой поверхности не является механической энергией. Вернее, для паров нарушается связь между их тепловой и механической энергиями, то есть она будет иная, чем для газов. Соответственно, пары находятся под действием иных силовых полей, например, электрических. В этом случае они составляют некоторую механическую смесь с атмосферой Земли, уровень механической энергии которой определяется газами состава её атмосферы. В силу этих причин мы ощущаем запахи даже очень тяжёлых веществ. В этом случае мы тоже имеем дело с парами, хотя мы их не видим. Практически пары любых веществ, даже воды, мы не можем видеть в силу их прозрачности. Видим мы в таких случаях не пары, а конденсат этих паров.

Теперь снова вернемся к атмосфере Венеры. Первый слой зоны энергетического прироста газов состоит из углекислого газа, потому что в условиях этого слоя и углекислый газ, и азотно-кислородная смесь будут находиться в состоянии газов. Эти газы в отношении друг к другу являются химически нейтральными. В этом случае разделение газов будет происходить по законам механики безынертной массы в зависимости от плотности этих газов. Количественно силовое различие газов выражается через объёмный вес. Азотно-кислородная смесь газов, как газ с меньшим объёмным весом, будет всплывать на поверхность углекислотного слоя. Вот таким способом сохраняется целостность углекислотного слоя газа, или первого слоя зоны энергетического прироста газов.

На высоте порядка 50 км, где давление равно приблизительно 1 ата и температура порядка 0°С, углекислота начинает конденсироваться, то есть на этой высоте углекислый газ утрачивает механическую энергию и переходит из газообразного в парообразное состояние в виде конденсата или просто пара, не способного воспринимать механическую энергию. Тогда эта высота становится граничной высотой толщины слоя углекислого газа. Над этим слоем будет располагаться слой азотно-кислородной смеси газов, который в этих условиях способен воспринимать механическую энергию. Энергетический уровень этого слоя будет определяться свойствами азотно-кислородной смеси. В этом случае углекислый газ первого слоя может находиться в виде паров во втором слое. Пары, как вы знаете, не изменяют уровень механической энергии. Ибо для них не существует объёмный вес, так как для этих паров верхний слой газов будет являться областью с нулевой механической энергией. Поэтому присутствие углекислоты в азотно-кислородном слое можно определить по степени насыщенности этого слоя парами углекислоты. Эту степень нельзя определить зависимостями механики безынертной массы. Она определяется либо химическими, либо физическими способами. Для нас главным здесь является то обстоятельство, что пары углекислоты не изменяют энергетического уровня.

В слое азотно-кислородной смеси газов до нижней границы подзоны постоянной температуры газов располагаются облака водяных паров согласно рис. 17. Облака водяных паров тоже не изменяют энергетического уровня слоя азотно-кислородной смеси газов зоны энергетического прироста газов. Об облаках водяных паров Венеры мы можем сказать, что они находятся в абсолютно одинаковых энергетических условиях с условиями водяных паров Земли с точки зрения законов механики безынертной массы, если судить по таблице рис. 17. Здесь наибольшее различие заключается лишь в том, что слой азотно-кислородной смеси сильно насыщен парами углекислоты. Коль мы не химики и не физики, то мы можем только предположить, что облака водяных паров Венеры ведут себя так же, как и облака водяных паров Земли. С той лишь разницей, что дождь, снег и град на Венере могут идти в виде газированной воды, снега или града. Поверхности Венеры ни дождь, ни снег, ни град, никогда не достигают, ибо они испаряются в слое углекислого газа и снова возвращаются в слой азотно-кислородной смеси. Подобная циркуляция воды и пара не подчиняется законам механики безынертной массы. Здесь мы можем отметить лишь то, что облака водяных паров Венеры располагаются в зоне застывшего адиабатического расширения азотно-кислородного слоя.

Слой азотно-кислородной смеси имеет подзону постоянных температур газов. Как мы знаем, эта подзона количественно определяет застывший расход массы через поверхность критических скоростей газов. По этой причине газ в этой подзоне имеет постоянную температуру. Для земной атмосферы она равна –40°С. Для венерианской атмосферы эта температура будет того же порядка, что и для земной атмосферы. Поверхность критических скоростей газов, которая характеризует критический расход газа зоны энергетического прироста газов, располагается на высоте 65 – 70 км от поверхности Венеры.

Затем идёт зона расширения газов. В этой зоне происходит уменьшение механической энергии газов атмосферы до минимума. Этот минимум характеризуется конденсацией или переходом газов в парообразное состояние. Этот минимум и определяет верхнюю границу зоны расширения газов, которая одновременно является конечной границей атмосферы Венеры. Она располагается на высоте порядка 100 км от поверхности Венеры. Зона расширения газов характерна тем, что в этой зоне могут располагаться пары углекислоты, так как пары не подчиняются законам механики безынертной массы, а более легкие газы (типа водорода, гелия) – нет. Ибо эта зона образована азотно-кислородной смесью, объёмный вес которой больше объёмного веса водорода, гелия, а разделение газов по объёмным весам в зоне расширения газов идёт в соответствии с законами механики безынертной массы.

Выше располагается зона нулевой энергии, или зона паров. Пары этой зоны не подчиняются законам механики безынертной массы. Пары этой зоны называют ещё ионизированными газами, а саму зону – ионосферой. Она характерна тем, что состав её может быть разнообразным по химическим веществам и элементам, так как он не определяется объёмным весом, а иными химическими и физическими свойствами. В ионосферу из зоны расширения газов поступают пары её газов, лишенные механической энергии, но обладающие определённой величиной тепловой энергии, которую они теряют в ионосфере. В результате чего происходит вынос тепловой энергии планеты в открытый космос. Таким способом организован отток тепловой энергии от планеты.

Для атмосферы планет принята классификация такого порядка как тропосфера, стратосфера и ионосфера. По нашей классификации зона тропосферы должна совпадать с зоной энергетического прироста газов, то есть граница тропосферы для атмосферы Венеры совпадает с поверхностью критических скоростей газов и находится на высоте 65 – 70 км над поверхностью Венеры. Границы стратосферы атмосферы Венеры совпадают с границами зоны расширения газов, то есть её нижняя граница совпадает с поверхностью критических скоростей газов, а верхняя находится на высоте порядка 100 км. По нашей классификации атмосфера Венеры будет состоять из тропосферы и стратосферы. Границы зоны нулевой энергии, или зоны паров, будут совпадать с границами ионосферы. Эта зона не подчиняется законам механики безынертной массы. По этой причине мы её не вносим в разряд атмосферы. Просто потому, чтобы показать, что она требует поисков своих законов. На рис. 17 покажем наши границы атмосферы Венеры.

Отметим, что в слое азотно-кислородной смеси тропосферы Венеры располагается слой водяных паров, толщиной 7 – 10 км. Этот слой почти не пропускает солнечную энергию к поверхности Венеры. По этой причине баланс тепловой энергии в слое углекислого газа тропосферы поддерживается тепловой энергией поверхности Венеры. Тепловой баланс всего остального объёма атмосферы Венеры поддерживается солнечной энергией. В этот объём входит слой азотно-кислородной смеси и зона расширения газов, или стратосфера. Отток же тепловой энергии происходит через объём ионосферы. Этим подчеркивается близкая схожесть атмосферы Венеры с атмосферой Земли.

В общем, используя данные показаний приборов автоматических станций «Венера» и «Маринер», которые мы брали из журнала «Наука и жизнь», мы можем только сопоставить атмосферы Земли и Венеры, что мы и сделали. Как бы косвенным путем подтвердили правильность положений нашей теории. Одновременно мы узнали много нового о планете Венера.

**Следующей планетой на нашем пути будет планета Земля.** Она отличается от ранее встретившихся нам планет тем, что в её гидросфере имеется плоский установившийся поток насосного типа. Поэтому она находится в состоянии динамического равновесия. Внешнее различие планет находящихся в состоянии статического равновесия с динамическим движением от планет, находящихся в состоянии динамического равновесия выражается в различии скоростей осевого вращения этих планет. Для планет находящихся в состоянии статического равновесия с динамическим движением тангенциальная скорость плоского установившегося потока гидросферы планеты не должна превышать постоянной скорости поля планетного тяготения. Для планеты Земля эта скорость равна $w$ = 3,132 м/сек. Выше мы определили, что тангенциальную скорость плоского установившегося течения гидросферы планеты нужно определять по движению экваториального диаметра планеты. Но Земля имеет земную кору, толщиной до 50-70 км. Поэтому мы должны были бы на эту величину уменьшить экваториальный диаметр. В связи с тем, что от полюса до полюса Земля вращается с одинаковой скоростью, то из условия равенства линейных тангенциальных скоростей при таком вращении мы должны были бы уменьшить экваториальный диаметр планеты, как диаметр внешней границы плоского установившегося течения, ещё на определённую величину. В общем, чтобы не гадать, где расположена внешняя граница плоского установившегося течения гидросферы планеты, мы просто полагаем, что она совпадает с внешней границей плоского установившегося потока той же гидросферы планеты. Тогда по условию равенства энергий (исходя, конечно, из средних величин физических характеристик массы планеты) внешняя граница плоского установившегося потока будет располагаться на половине экваториального диаметра планеты. Экваториальный диаметр Земли имеет величину порядка 12400 км. Тогда половина её диаметра будет равна 6200 км. Этот диаметр будет средним внешним диаметром внешней границы плоского установившегося потока гидросферы Земли. Земля делает один оборот за 24 часа. За это время линейная скорость пробегает длину окружности с диаметром 6200 км. Тогда длина $L$ этой окружности будет равна:

$$L = \pi d = 3{,}14 \cdot 6\,200 = 19468\ (км).$$

В этом случае тангенциальная линейная скорость в час будет равна:

$$W_{tg} = L:24 = 19468:24 = 811\ (км/час).$$

Тангенциальная скорость за одну секунду будет равна:

$$W_{tg} = 811000(м):3600\ (сек) = 225\ (м/сек).$$

Как видите, тангенциальная скорость плоского установившегося потока во много раз превышает постоянную скорость поля планетного тяготения, которая равна для Земли $w$=3,132 м/сек. Это подтверждает, что в гидросфере планеты Земля размещается плоский установившийся поток насосного типа, а сама она находится в состоянии динамического равновесия. Если же вы подобные вычисления сделаете для Меркурия и Венеры, то вы убедитесь, что тангенциальные скорости их плоских установившихся течений не превышают постоянной скорости их полей планетного тяготения.

Нам известна теперь средняя величина тангенциальной скорости плоского установившегося потока гидросферы Земли. Ради интереса попробуем вычислить среднюю величину сил давления на единицу площади сечения потока в тангенциальном направлении. Она будет равна:

$$P_{tg} = \rho W_{tg}^2 .$$

Для Земли средняя плотность принята равной $\rho$ = 5,5 г/см$^3$. Для одного кубического метра она будет равна $\rho$ = 5,5·10$^3$ кг/м$^3$. Тогда тангенциальная сила давления единицы площади будет равна:

$$P_{tg} = 5{,}5 \cdot 10^3 \cdot 225^2 = 2{,}75 \cdot 10^8\ (кг/м^2).$$

Цифра получилась очень значительная. Как вы помните из расчёта центробежных насосов, эти силы создаются вращающейся лопаткой колеса насоса. Если эту величину тангенциальных сил давления отнести к единице объёма потока, то мы получим величину полной энергии единицы объёма потока. Теперь, если мы эту величину умножим на полный объём плоского установившегося потока гидросферы, то получим полную энергию этого потока. Мы получили первое количественное представление о плоском установившемся потоке гидросферы Земли.

В центре планеты, вернее, во входном потоке её плоского установившегося потока располагается источник тепловой энергии. Ибо он является одним из основных условий, которое необходимо для существования плоского установившегося потока в гидросфере планеты. Выше мы показали, что плоский установившийся поток гидросферы поглощает определённую часть тепловой энергии источника тепла, расположенного в центре Земли. Тем самым он снижает тепловую энергию внутренней массы планеты. Мы показали это изменение через уменьшение плотности массы в центральной области гидросферы планеты. Ибо это условие следует из законов механики безынертной массы, которые связывают механическую энергию с тепловой через изменение плотности массы. Помимо плотности веществ связанной с изменением давления существует просто различная плотность веществ, которая связана с их химическим составом. В соответствии с законами механики безынертной массы при равных условиях движения жидкие и газообразные вещества с большей плотностью будут воспринимать большую силу со стороны силового поля, чем вещества с меньшей плотностью. По этой причине вещества с большей плотностью будут в первую очередь выноситься из центра планеты её плоским установившимся потоком за пределы внешней границы этого потока. Это условие способствует накоплению более плотных химических веществ за пределами плоского установившегося потока гидросферы Земли.

В связи с тем, что часть тепловой энергии источника тепла планеты поглощается её плоским установившимся потоком, то оставшейся части тепловой энергии не хватает на содержание всей массы гидросферы Земли в жидком состоянии. Недостаток тепловой энергии привёл к образованию твёрдой земной коры на поверхности гидросферы. Ибо источник тепловой энергии Земли по своей мощности не меньше, например, источника тепловой энергии Венеры.

Земная кора своей толщиной не нарушает полной механической энергии планеты. Это связано с тем, что земная кора плавает в расплавленной магме гидросферы планеты в соответствии с законом Архимеда. Это значит, что земная кора, образовавшись из магмы гидросферы планеты, как бы вобрала в себя часть массы гидросферы Земли. Поэтому плавание земной коры будет возможно лишь в том случае, если её плотность будет меньше, чем плотность магмы гидросферы. В связи с тем, что вес земной коры остаётся неизменным, то и вытесненный ею объём жидкости тоже остаётся неизменным. Это значит, что уровень магмы гидросферы Земли остался неизменным после образования земной коры. Земная кора возвысилась над уровнем магмы, или гидросферы планеты, лишь на величину, которая зависит от разностей плотностей земной коры и магмы. Например, мы знаем, что лед в воде погружается на две трети своего объёма. Пробив во льду лунку, мы можем определить истинный уровень воды в водоеме. Для Земли подобными лунками могут служить кратеры действующих вулканов. Как мы знаем, плавающий лед не нарушает энергетического баланса водоёма, то есть давление на глубине этих водоемов остаётся неизменным. Так и земная кора не нарушает энергетического баланса гидросферы Земли, коль она не изменяет граничной поверхности гидросферы для расплавленной магмы.

Отметим, что истинный диаметр гидросферы Земли определяется уровнем её магмы, а не поверхностью её Мирового океана, как мы привыкли считать. Это существенное замечание, которое приведет к изменению некоторых физических характеристик Земли в нашем представлении. Например, мы будем вынуждены уменьшить диаметр Земли. В связи с этим изменится её средняя плотность. Она должна возрасти против известной нам цифры. Всё это можно выявить дополнительными исследованиями.

Вот то всё новое, что мы смогли узнать о гидросфере нашей планеты Земля, используя законы механики безынертной массы.

Видите ли, какая тут зависимость наблюдается. Всё то новое, что мы с вами получили о строении гидросферы Земли, нельзя получить простыми исследованиями связанными с непосредственными наблюдениями и измерениями картин планетных явлений. Ибо в этих практических исследованиях выделяются в первую очередь сами измеряемые и наблюдаемые величины явлений, относительно которых в последующем пытаются оценить само явление. По этой причине упускают ту общность, которая, например, определяется для планет законами механики безынертной массы. Это значит, что наблюдения нашего путешествия, которые мы ведём с точки зрения законов механики безынертной массы, по своей необходимой значимости должны быть первичными, а практические исследования – вторичными. Ибо только после наших наблюдений нового практические исследования могут носить целенаправленный характер. Только такие целенаправленные практические исследования впоследствии могут дать чёткие количественные границы нашему новому. Если бы мы поступили наоборот, то есть поставили первичными практические исследования, а вторичными наши наблюдения планет с точки зрения механики безынертной массы, то мы бы с вами не заметили ничего нового в нашем путешествии. Мы просто топтались бы на том же уровне знаний, которые известны всем и которые до сих пор никто не может расширить. Такое положение может продолжаться очень долго во времени, то есть в течение десятков, а то и сотен лет. Поэтому относительно выявления нового в строении планет наши наблюдения являются первейшей необходимостью, без которой нельзя выявить это новое. Ибо до подобного момента всякие практические исследования явлений природы считаются как научные исследования вообще, а конкретность и целенаправленность они приобретают только после выявления нужной общности, выраженной, например, как в законах механики безынертной массы.

Данное замечание, или отступление, сделано для того, чтобы вы отнеслись с должным вниманием ко всему новому, что вы наблюдаете в вашем путешествии от планеты к планете, которое не должно оставлять вас равнодушными. Если оно возбудит в вас гордость и восхищение, которое затем будет подкреплено соответствующей вашей деятельностью, то вы получите чисто человеческую удовлетворенность. Если в вас всё это вызовет напыщенное чванство, которое неизвестно для чего нужно даже индюку, или простое равнодушие, то знайте, что вы в этом случае ещё не доросли до человеческого сознания и вам необходимо будет как можно скорее ликвидировать этот пробел. Наша работа поможет вам ликвидировать его и дорасти до человеческого сознания лишь в том случае, если с вашей стороны будут приложены максимум усилий и труда, которые необходимы для формирования в вашем сознании полной ясности по всем положениям механики безынертной массы.

**Перейдем к атмосфере Земли**. Она представляет собой однослойную азотно-кислородную смесь. Зона энергетического прироста газов, или тропосфера, атмосферы простирается до высоты 25 км. Подзона постоянных температур газов занимает примерно половину высоты тропосферы. Ограничивается она на высоте 25 км поверхностью критических скоростей газов, которая одновременно является границей тропосферы (см. рис. 15). Температура газа в этой подзоне постоянна и равна –40°С. Выше тропосферы располагается зона расширения газов, или стратосфера, атмосферы Земли. В этой зоне температура газа растет по причине убыли механической энергии. Верхняя граница стратосферы располагается на высоте порядка 100 км. Верхняя граница стратосферы одновременно служит границей атмосферы Земли. Выше атмосферы располагается зона нулевой энергии газов, или зона паров, которая в общепринятой терминологии называется ионосферой. Через эту зону происходит вынос тепловой энергии планеты в открытый космос. В общем, мы здесь несколько изменили общепринятые границы тропосферы, стратосферы и самой атмосферы Земли в соответствии с законами механики безынертной массы.

Далее рассмотрим особенности земной атмосферы. По нашей модели нижняя граница атмосферы совпадает с границей гидросферы планеты. Для атмосферы Земли мы взяли нижнюю её поверхность от поверхности Мирового океана, которая принята для начала отсчётов различных поверхностных уровней. При разборе гидросферы Земли мы установили, что уровень Мирового океана располагается приблизительно на 50 км выше границы гидросферы из-за того, что земная кора плавает в магме гидросферы. Эти 50 км земной коры являются прослойкой между границами гидросферы и атмосферы Земли. Это не значит, что нет контакта между атмосферой и гидросферой. Если бы не было такого контакта, то химический состав атмосферы со временем должен был бы измениться. Ведь современная промышленность и техника поглощает очень большое количество кислорода из атмосферы Земли. Поэтому можно сказать, что атмосфера является дыханием земной гидросферы.

Главной особенностью этой прослойки является то обстоятельство, что она служит хорошей теплозащитой между гидросферой, которая имеет сравнительно высокие температуры, и атмосферой, которая является потребителем тепловой энергии. Вот эта прослойка как бы отгораживает атмосферу Земли от тепловой энергии её гидросферы. Трудно судить о количестве тепла, поступающего из гидросферы Земли в её атмосферу не имея соответствующих цифр, но всё равно мы можем сказать с большой уверенностью, что этот приток тепла очень незначителен. Ибо на материках Земли существуют зоны вечной мерзлоты и материковые льды, которые имеют очень низкий температурный режим. Поэтому мы будем считать, что баланс тепловой энергии земной атмосферы поддерживается за счёт тепловой энергии Солнца. Это значит, что Солнце не только светит, но и греет. Тем самым оно полностью поддерживает биологическую жизнь на Земле.

Следующей особенностью земной атмосферы является то обстоятельство, что экваториальный радиус гидросферы Земли на 21 км больше полярного. Как вы знаете, это отклонение от шарообразной формы гидросферы Земли определяется повышенной тепловой энергией во входном потоке её плоского установившегося потока насосного типа. Форма земной атмосферы шарообразна, ибо она полностью образована земным полем тяготения. По этой причине толщина атмосферного слоя Земли в области полюсов должна быть больше, чем на экваторе. Это значит, что при одинаковой распределённости тепловой энергии по всей толще земной атмосферы механическая энергия в области полюсов планеты будет по величине больше механической энергии атмосферы в экваториальной зоне.

В результате этих двух особенностей земная атмосфера в области полюсов имеет максимальную механическую энергию и минимум тепловой, а в экваториальной зоне она имеет максимум тепловой энергии и минимум механической. Это неравенство энергий является причиной различных воздушных течений в земной атмосфере. Это неравенство может усиливаться неравномерным прогревом поверхности Земли, что является причиной различных ураганных движений воздуха. Обычно различают струйное и циклонное движение воздуха в атмосфере. Струйное течение вам понятно, а циклонное представляет собой ни что иное, как движение плоского установившегося потока атмосферы Земли. Наиболее сильное перемещение воздушных масс наблюдается в период смены времён года. Ибо это связано с изменением теплового режима целого полушария Земли. Если бы Земля имела идеальную шаровую форму, то средняя температура в области полюсов повышалась бы, а в экваториальной зоне – понижалась бы. Ибо в этом случае теплый воздух с экватора Земли беспрепятственно проникал бы в области полюсов, так как повышенная механическая энергия полюсных зон атмосферы Земли не препятствовала бы распространению повышенной тепловой энергии экваториальных зон. Мы бы просто имели более усредненные температуры в полушариях Земли с несколько уменьшенными максимальными и минимальными температурами в экваториальных и полюсных зонах. При этом все перемещения воздушных масс в земной атмосфере имеют прямую связь с зонным энергетическим распределением по высоте атмосферы, то есть с её тропосферой и стратосферой, не исключая и ионосферу. Ибо эти зоны определяют форму движения воздушных масс, ограничивая её по высоте и по энергетическому уровню, связанного с высотой.

Вам кажется, что мы начали повторяться. Вот умный синоптик этого не скажет. Он просто придёт в сильный восторг от нашего описания атмосферы Земли. Естественно, его внимание привлечет не сам стиль нашего описания, который далек от лучших образцов, а от тех возможностей, которые ему даёт наша работа как предсказателю погоды. Вот что привлечёт его внимание. Ибо каждому специалисту приходится переворачивать множество различных описаний, чтобы найти хотя бы то немногое, которое помогает ему сделать совершеннее свой труд. В большинстве случаев он в этих описаниях не находит для себя ни грамма того, чего бы он не знал и чего бы он мог использовать в своей трудовой деятельности. По этой причине ему часто приходится испытывать разочарование и огорчение за напрасно потраченное время. В нашей работе он находит сразу очень многое для себя полезного, а отсюда вытекает его радость и восторг. Затем, чтобы продлить своё хорошее настроение, наш умный синоптик постарается сразу же пустить в дело полученные знания. В этом случае его начинает интересовать количественная сторона дела, то есть цифры, которые он мог бы сопоставить с практическими данными.

Для этого он начал бы действовать в таком плане. Сначала он уяснил бы для себя, что всякая масса, в каком бы она состоянии не находилась: в твёрдом, жидком или газообразном в зависимости от мощности источника тепловой энергии и от его полного количества тепла, через некоторое время насыщается до определённого предела тепловой энергией, после которого количество тепла, притекающего в единицу времени к телу определённой массы и количество тепла утекающего от этого тела в единицу времени будут равны по величине друг другу независимо от времени. Конечно, при условии, что источник тепловой энергии не изменяет в течение всего этого времени своего режима работы. Предел насыщения тепловой энергией каждого тела определяется по величине температур. После достижения предела насыщения тепловой энергией температура тела перестаёт изменяться и устанавливается на уровне в соответствии с эти пределом. Таким телом является атмосфера Земли. Своим газообразным состоянием она обязана только тепловой энергии. Вот таким способом синоптик определил бы общие условия для термодинамического состояния атмосферы, то есть он бы получил атмосферу Земли как термодинамическое тело.

Далее из нашего описания атмосферы он понял бы, что источником тепловой энергии атмосферы является Солнце, а утечки тепловой энергии происходят через ионосферу. Что атмосфера Земли находится в термодинамическом равновесии, то есть приток и отток её тепловой энергии равны между собой по величине. При этом приток тепловой энергии полностью зависит от источника тепловой энергии, а отток – от термодинамического состояния атмосферы. Составив такие условия, умный и грамотный синоптик сразу увидит, что для атмосферы Земли он может пользоваться зависимостями термодинамики в полной мере. Поэтому он, как специалист своего дела, никогда не сошлётся на то обстоятельство, что, мол, мы не дали ему количественных величин тепловой энергии Солнца, что мы не дали ему количественных характеристик атмосферы и т.д. Ибо он знает, чтобы дать подобные количественные величины, нам бы всё равно пришлось обратиться к нему же за этими величинами. Ведь он как специалист-синоптик производит постоянный замер этих величин с помощью специальных приборов и установок, а нам приходится пользоваться всего-навсего количественными данными из учебника астрономии. Зависимостей мы здесь не приводим лишь потому, что они есть в термодинамике и в механике безынертной массы, зависимости и законы которой мы приводим в данной работе. Умный и грамотный синоптик приходит в восторг от нашего описания потому, что он видит в этом описании возможности к применению своих количественных величин, которые он получает с помощью замеров как специалист-синоптик.

Посмотрим, что будет делать умный специалист-синоптик дальше. Дальше он возьмёт нижний слой атмосферы Земли, который называется тропосферой. Самый нижний слой тропосферы, который находится у поверхности Земли и ограничивается нижней границей подзоны постоянных температур. В этом слое зафиксировано застывшее расширение газов в соответствии с адиабатическим процессом. Поэтому синоптик воспользуется для этого слоя зависимостями адиабатического процесса. Об активном процессе адиабатического расширения газов говорят, что уменьшение температур в потоке происходит за счёт расширения газов. В застывшем адиабатическом расширении газов атмосферы снижение температур происходит за счёт уменьшения тепловой энергоёмкости по высоте слоя, а тепловая энергоёмкость уменьшается за счёт уменьшения плотности газов, которая происходит в соответствии с законами и зависимостями адиабатического процесса термодинамики.

Для подзоны постоянных температур газов синоптик примет тепловую энергоёмкость постоянной величиной. Постоянство температур этой подзоны указывает на постоянство плотности газа. Поэтому механическая энергия этой подзоны определяется как для жидкости с постоянной плотностью в зависимости от высоты или толщины слоя по зависимостям механики безынертной массы, а тепловая энергоёмкость определяется по зависимостям термодинамики как количество тепла, которое необходимо для превращения жидкой массы в газообразную до соответствующего температурного уровня подзоны постоянных температур газов. В этой подзоне газ ведёт себя как несжимаемая жидкость. С точки зрения застывшего механического движения газов она определяется как застывший критический расход массы газа тропосферы Земли. Поэтому верхней границей подзоны постоянных температур газов служит поверхность критических скоростей газов. Критическая скорость газов этой поверхности равна постоянной скорости поля земного тяготения, то есть $w$ = 3,132 м/сек.

Для зоны расширения газов, или стратосферы Земли, синоптик тоже воспользовался бы зависимостями адиабатического процесса термодинамики. Здесь тоже происходит уменьшение тепловой энергии по высоте, или толщине стратосферы, в зависимости от уменьшения плотности газов по высоте стратосферы, которое определяется зависимостями адиабатического процесса. Различие здесь с расширением газов тропосферы Земли будет заключаться в том, что тепловая энергия в тропосфере сохраняется или поддерживается на соответствующем уровне за счёт механической энергии, а в стратосфере происходит как бы преобразование или полный переход механической энергии в тепловую. По этой причине в тропосфере Земли происходит уменьшение температур по высоте, а в стратосфере – увеличение температур по высоте. Вводя новое понятие о тепловой энергоёмкости газов в зависимости от их плотности, мы тем самым определим, что механическое движение под действием сил поля земного тяготения и движение расширения зависящее от тепловой энергоемкости газов атмосферы, являются застывшими движениями, а тепловое движение, или поток тепловой энергии в атмосфере, является не застывшим, то есть происходит непрерывный переток тепла в атмосфере Земли от поверхности к периферийной границе. По этой причине активный тепловой поток как бы сохраняется в объёме массы газов тропосферы, а в объёме массы газов стратосферы происходит освобождение теплового потока, которое мы определили выше как переход механической энергии в тепловую. Именно так он определяется зависимостями механики безынертной массы и термодинамики. По этой причине изменение плотности газа в атмосфере и стратосфере Земли определяется зависимостями адиабатического процесса, но в тропосфере происходит уменьшение температур, а в стратосфере – увеличение температур газов.

Верхнюю границу стратосферы, которая является одновременно границей атмосферы Земли, мы определили минимальной энергией конденсации газов. Ибо по зависимостям адиабатического процесса определяя уменьшение плотности газа по высоте стратосферы мы одновременно определяем её как уменьшение температуры по той же высоте стратосферы. По этим результатам мы, в конечном итоге, придем к минимальной энергии конденсации газов стратосферы. По этой причине мы в любом случае называем подобные границы границами конденсации. В одних случаях они совпадают с истинными границами конденсации, например, слой углекислого газа тропосферы Венеры, а по отношению к стратосфере Земли это определение будет формальным. Просто эта граница определяет утрату механической энергии на этой границе. Выше этой границы газы переходят в ионизированные газы, или пары газов, состояние которых определяется иными законами.

По этой причине синоптику придётся интересоваться состоянием ионосферы Земли у других специалистов типа физиков.

Сейчас мы рассматриваем как бы план действий умного синоптика. Вот он прочитал наше описание атмосферы Земли или другой какой-либо планеты. Затем отвлёкся от него и начинает составлять план своих действий для решения своих специальных задач. Делая наше описание его действий, мы как бы следим за его мыслью. Вышеизложенным описанием он проверял, во всех ли зонах атмосферы он может воспользоваться зависимостями и где их можно взять для конкретного количественного решения. Всё это он делал быстро в уме. Коль он грамотный специалист, то он знаком с другими науками, которые необходимы для его специальности.

После того, как он выяснил, что для всех зон атмосферы есть количественные зависимости термодинамики и механики безынертной массы, перед ним встаёт новый вопрос, который заключается в том, чтобы найти начало отсчёта для этих зависимостей, или сделать привязку к конкретным условиям. Поэтому дальше он будет мыслить следующим образом. Температурные режимы меняются, вместе с ними меняются и плотности газов атмосферы. В то же время давление на поверхности Земли почти не изменяется из-за того, что количество воздуха в атмосфере Земли есть постоянная величина. Плюс ко всему существует суточное вращение Земли, что приводит к периодическому прогреву участков Земли Солнцем. Существуют также сезонные и годовые изменения температурных режимов земной атмосферы. Помимо всего, в земной атмосфере содержится очень много водяных паров и т.д. Вот сколько мы выявили различных условий и все их надо совместить, чтобы получить правильное количественное решение. Всё это сделать под силу только специалисту-синоптику. Все эти условия он тоже будет совмещать в определённом порядке.

Начнёт он его с построения энергетически равновесной модели земной атмосферы. Начальной основой для этого он выберет поверхность Земли в виде формы поверхности Мирового океана, которая принята для всех современных земных отсчётов. Одновременно она является нижней границей земной атмосферы. Второй базовой поверхностью этой модели будет поверхность критических скоростей газов. Ибо критическая скорость газов атмосферы по всей поверхности одинакова и равна $w$ = 3,132 м/сек. Она зависит от поля земного тяготения. По этой причине поверхность критических скоростей газов должна иметь идеальную шаровую форму. Синоптик может измерить высоту этой поверхности, а затем определить её радиус от центра Земли, который будет одинаковым как для экваториальных, так и полюсных зон Земли. После чего он наносит на свою модель вторую базовую поверхность. Выше он установил, что в экваториальных зонах существует максимум по температурам, а в полюсных зонах – максимум по механической энергии у поверхности Земли. По этой причине должна существовать определённая циркуляция земной атмосферы. Синоптику же надо получить энергетически уравновешенную атмосферу, в которой бы отсутствовала эта циркуляция. Для этого он должен увидеть способы для равновесной компенсации этой циркуляции атмосферы. Таким единственным способом является то обстоятельство, что через поверхность критических скоростей газов атмосферные газы должны двигаться с постоянной скоростью равной $w$ = 3,132 м/сек. По этой причине объёмный расход газов будет по всей поверхности критических скоростей одинаков по величине, а по массе – различным. Ибо объёмный расход не зависит от температуры, а массовый зависит, так как от температуры зависит плотность газов. По этой причине в области экватора, где подзона постоянных температур тропосферы имеет более высокие температуры, чем в области полюсов, расход массы газа будет меньше, чем в области полюсов Земли через поверхность критических скоростей газов. Коль мы рассматриваем застывшее движение расширения газов атмосферы, то повышенные и пониженные массовые расходы через поверхность равных скоростей выразятся в толщине, или в высоте, подзоны постоянных температур газов. Это значит, что в области экватора эта подзона будет иметь большую толщину, чем в области полюсов Земли. Поэтому из-за пониженного расхода массы в области экватора в подзоне постоянных температур газов будет накапливаться большая масса газа, а в области полюсов из-за повышенного расхода массы будет накапливаться меньшая масса газа. Непосредственно само давление на поверхности Земли зависит в первую очередь от количества массы газа в подзоне постоянных температур газа. Чем больше эта масса, тем большее давление должен иметь нижний слой тропосферы, чтобы держать эту массу. Ведь сила и энергия механического застывшего движения находится в прямой зависимости от количества массы. Поэтому за счёт изменения количества массы газа в подзоне постоянных температур газа тропосферы вышеуказанным способом можно получить равенство механических энергий в области экватора и в области полюсов Земли.

Получив равенство механических энергий на экваторе и на полюсах Земли, мы дальше распространяем эту величину механической энергии и на остальную поверхность Земли, расположенную между экватором и полюсами в соответствии с их условиями. После чего синоптик получит модель атмосферы Земли уравновешенную по механической энергии. Солнечная тепловая энергия имеет свой максимум в области экватора и минимум – в области полюсов. Вот эта энергетически уравновешенная модель атмосферы Земли будет являться началом отсчёта для изменяющихся энергетических режимов непосредственно самой атмосферы Земли. Всякие или любые отклонения температуры или давления в модельной атмосфере приведут к нарушению энергетического равновесия атмосферы, что повлечёт за собой определённые формы движения воздуха атмосферы. По этой причине уравновешенную по механической энергии модель атмосферы необходимо принять за начало отсчётов, то есть начальной конкретной величиной для различных зависимостей.

Составляя эту модель синоптик будет исходить из реальных условий. Для него реальные условия будут таковы, что для экваториальной зоны поступление тепловой энергии Солнца не зависит от времени года, а в области полюсов и на остальной поверхности Земли оно зависти от времени года. На полюсах Земли полгода бывает день, полгода – ночь, то есть полгода – лето, полгода – зима. В зависимости от времени года меняется температура в области полюсов. Поэтому в летний сезон соответственно поднимется температура и в подзоне постоянных температур газа тропосферы в области полюсов. В результате чего произойдет увеличение воздушной массы этой подзоны по сравнению с зимним сезоном полюсной области, поскольку мы знаем, что повышенные температуры способствует накапливанию массы воздуха в подзоне постоянных температур газа. По этой причине в области полюсов механическая энергия тропосферы бывает в летний сезон большей, чем в зимний. Тогда перед синоптиком сразу встанет вопрос, какую величину этих энергий - летнюю или зимнюю - необходимо ему взять для составления модели атмосферы уравновешенной по механической энергии.

Далее его задача усложняется тем, что на полюсах Земли в одно и то же время бывают разные времена года. Если на северном полюсе зима, то на южном – лето или наоборот. При этом подобное различие касается обоих полушарий Земли. Вот эти условия усложняют модель атмосферы уравновешенной по механической энергии. Поэтому синоптику придётся хорошо подумать, прежде чем он получит необходимую для себя и правильную модель, которая бы вписывалась в годовые колебания механической энергии по полушариям Земли. Мы не будем мешать ему в этом. Ибо с этой задачей он справится лучше нас. Мы лишь поинтересуемся для общего обзора тем, какие перемещения воздушных масс вызывают сезонные изменения энергий на полушариях Земли.

Для этого мы опять должны вспомнить, что количество воздушной массы в атмосфере Земли есть величина постоянная. В то же время повышенные температуры приводят к накоплению массы воздуха в тропосфере Земли. Повышенные температуры тропосферы будут на полушарии Земли, где стоит лето, а пониженные – где зима. Отсюда следует, что на полушарии Земли в период летнего сезона количество воздушной массы бывает большим, чем в зимний сезон. Поэтому общая воздушная масса атмосферы между полушариями Земли делится не поровну, а большая её часть находится в полушарии с летним сезоном, меньшая – в полушарии с зимним сезоном. Это значит, что при замене времен года на полушариях Земли одновременно происходит переток излишков воздушной массы из одного полушария в другое. Мы наблюдаем четыре времени года: зиму, лето, весну и осень. Летом на нашем полушарии накапливается больше воздушной массы, чем зимой. Весной и осенью происходит приток или отток излишков воздушной массы из нашего полушария. В эти периоды происходит переход атмосферы полушария Земли на новый энергетический уровень. По этой причине мы наблюдаем неустойчивую погоду в осенний и весенний периоды, с сильными ветрами и с повышенным количеством выпадающих осадков. В летние и зимние периоды погода бывает сравнительно устойчивая. Это значит, что в эти периоды атмосфера Земли стремится сохранить сезонное равновесие по механической энергии.

В области экватора Земли наблюдается всего два времени года – это период дождей и период сухой погоды. Период дождей в экваториальной области Земли будет связан с весенне-осенними периодами в её полушариях. Ибо в эти периоды происходит переток излишков массы атмосферы из полушарие в полушарие, который уносит из экваториальной области более теплый воздух и замещает его более холодным воздухом. В результате чего происходит быстрая конденсация водяных паров в этой области атмосферы. Что сопровождается обильным выпадением осадков. Только это может быть причиной образования периода дождей в экваториальной области Земли, так как поступление количества солнечной тепловой энергии во все времена года остаётся неизменным в этой области. В летние и зимние периоды полушарий Земли переток воздушных масс в области экватора отсутствует. Поэтому воздух прогревается до соответствующих температур и испарившаяся влага почти не конденсируется, что приводит к сравнительно небольшому количеству выпадения осадков в экваториальной области.

Следующие энергетические изменения атмосферы будут связаны с осевым, или суточным, вращением Земли и с различием физических условий её поверхности. При суточном вращении Земли в различное время суток количество тепловой энергии Солнца, поступающее на поверхность Земли, будет различным. В полуденные часы оно достигает максимума, а в ночные часы оно отсутствует. Тем самым создаётся суточный энергетический пик для поступления тепловой энергии Солнца, который бы приводил к резким изменениям температуры, если бы не существовало определённых физических условий на поверхности Земли.

Эти условия связаны с тем, что атмосфера Земли прозрачна для тепловой энергии Солнца. Разогрев атмосферы происходит от поверхности Земли, где лучистая солнечная энергия преобразуется в тепловую. Поэтому в первую очередь происходит разогрев земной поверхности, а от неё начинает разогреваться атмосфера. Одновременно происходит испарение воды с поверхности, которое тоже потребляет определённое количество тепловой энергии Солнца. Прогрев поверхности Земли и испарение воды сглаживают дневной максимум поступления тепловой энергии тем, что, как бы аккумулируя часть тепловой энергии, они тем самым уменьшают поток тепла в атмосферу. Ночью же, когда приток солнечной энергии отсутствует, разогретая поверхность Земли и испаренная вода продолжают наполнять атмосферу тепловой энергией, которую они накопили за световой день. В результате чего происходит не пиковая подача солнечного тепла в атмосферу, а сравнительно равномерная подача этого тепла в течение всех суток. Что приводит к сравнительно небольшому диапазону колебаний температур в течение суток. При этом подобные суточные колебания охватывают нижние слои тропосферы, расположенные в непосредственной близости от поверхности Земли. Подзона постоянных температур газов тропосферы по этой причине не испытывает, или не ощущает, подобных температурных колебаний. На эти условия ещё накладываются местные различия физических условий поверхности Земли. Например, песчаные, скалистые, обезвоженные участки поверхности прогреваются быстро и до более высоких температур. По этой причине прилежащие к этой поверхности слои воздуха прогреваются до более высоких температур. Ночью же, когда нет поступления солнечной энергии, скалистые и песчаные участки поверхности быстро отдают своё тепло воздуху. Поэтому на этих участках образуются резкие колебания температуры воздуха в пределах суток. На других участках поверхности Земли, где рыхлые и влажные почвы или на водных просторах, прогрев идёт медленнее. Прилежащие атмосферные слои воздуха на этих участках прогреваются за счёт испаренной влаги. В ночное время она тоже пополняет запасы тепловой энергии атмосферных слоёв воздуха. Тем самым как бы сглаживается суточный температурный пик.

В результате местных физических различий поверхности Земли образуются локальные энергетические различия атмосферных слоев воздуха над этими участками. Что приводит к местному различию уровней механической энергии в атмосфере, которое в свою очередь приводит в движение местные массы воздуха. Если это различие мало, то мы наблюдаем перемещение масс воздуха в виде слабого ветра, если это различие велико, то оно выражается в виде сильного или даже ураганного ветра. Вот таким способом влияют местные условия поверхности Земли и её суточное вращение на энергетический уровень атмосферы.

Отметим, что абсолютно любое движение воздушных масс существует в двух видах движения – в струйном и вихревом. Струйное движение воздуха относится к разряду установившегося вида движения жидкостей и газов. Вихревое движение воздушных масс относится к разряду плоского установившегося вида движения жидкостей и газов. Эти виды движения воздушных масс наблюдаются как при сезонных изменениях уровней энергий атмосферы, так и при местных. Просто струйное движение воздушных масс образуется между соседними участками атмосферы в горизонтальной плоскости, где по каким-либо причинам образуются различные уровни механической энергии. Вихревое движение воздушных масс атмосферы, типа тайфунов и смерчей, образуется из-за различия энергетических уровней по высоте атмосферы. Причиной подобных потоков обычно являются местные физические условия, которые приводят к энергетическому различию по высоте атмосферы. Видимый вертикальный жгут подобных потоков есть ни что иное, как внутренняя граничная поверхность плоского установившегося потока, которая образована окружностями равных скоростей. Высота этого потока будет граничить с теми слоями атмосферы, куда происходит переток воздушных масс. По-видимому, верхней границей тайфуна является нижняя граница стратосферы или поверхность критических скоростей газов, которая препятствует энергетическому равновесию. Всё это можно уточнить непосредственными замерами.

Как мы уже отметили, что подобные потоки относятся к разряду плоского установившегося вида движения. Остаётся добавить, что этот поток будет относиться к турбинному типу. Ибо здесь воздушные массы поступают в поток через его вертикальную периферийную границу и выходят через его внутреннюю граничную поверхность. В объёме этой поверхности они преобразуются в струйный поток, который затем поступает в слой воздуха с пониженным уровнем энергии в виде вертикальной воздушной струи. Воспользовавшись зависимостями механики безынертной массы и необходимыми натурными замерами, мы сможем для этого потока определить его характеристики, включая сюда и различие по энергетическому уровню слоев атмосферы. Количественные характеристики смерчей и тайфунов в будущем позволят найти необходимые меры борьбы с подобными явлениями, которые относятся к разряду стихийных бедствий. Циклоны и антициклоны относятся тоже к вихревому движению воздушных масс, но с той разницей, что они образуют плоские установившиеся течения воздушных масс. Ибо они не нарушают энергетического уровня механической энергии атмосферы. Движение воздушных масс подобного течения происходит за счёт частичного перехода полной механической энергии в кинетическую. Этот переход тоже образуется за счёт различия по тепловой энергии между объёмом воздуха плоского установившегося течения и окружающим его воздухом атмосферы. В этом случае происходит спокойный теплообмен между областями атмосферы. Плоские установившиеся течения, или циклоны, тоже определяются зависимостями механики безынертной массы.

Мы составили примерный план действий синоптика, который, полагаем, понял наше описание атмосферы Земли как специалист своего дела. После составления подобного плана действий он бы понял, что он может пользоваться существующими количественными зависимостями механики безынертной массы и термодинамики. В этом случае точность его количественных расчётов атмосферы полностью зависела бы от точности натурных измерений. Составив расчёты, он, скажем, получил бы точный прогноз погоды на одни сутки. Затем, усовершенствовав свои замеры, через некоторое время он бы получал точные прогнозы уже на неделю. Так, совершенствуя свои замеры и расчёты, он будет с течением времени получать точные прогнозы на месяц, на полгода и т.д. Вот когда он доведёт свои расчёты и замеры до полного количественного совершенства, то он начнёт строить планы преобразования атмосферы Земли. Планы такого специалиста-синоптика, в отличие от ныне существующих подобных планов, будут иметь реальную основу, которая будет подкреплена реальными человеческими возможностями и человеческой необходимостью.

<…>[4]

**[*Строение атмосферы и гидросферы планет-гигантов*]**

Далее мы совершим осмотр планет-гигантов: Юпитера, Сатурна, Урана и Нептуна. Каждая из этих планет во много раз больше нашей Земли. Они имеют быстрое суточное вращение порядка 10 часов и сильно сжаты у полюсов. Для таких больших планет, как Юпитер и Сатурн, разница между экваториальными и полярными радиусами достигает 6000 км.

О гидросфере этих планет мы можем сказать, что она находится в состоянии динамического равновесия. Что в гидросферах этих планет размещается плоский установившийся поток насосного типа. В центре планет, или во входном потоке плоского установившегося потока, размещается источник тепловой энергии. Отметим ещё, что тепловое энерговыделение на единицу теплового источника будет больше у планет, у которых полюсное сжатие объёма планеты достигает большей величины. Поскольку причиной увеличения объёма входного потока плоского установившегося потока является общее количество тепловой энергии, выделяемой в единицу времени источником тепловой энергии планеты. Большое полюсное сжатие Юпитера и Сатурна указывает на то обстоятельство, что их гидросфера находится на предельной границе динамического равновесия.

Как пишут, что «по спектральному наблюдению атмосферы планет-гигантов содержат в основном молекулярный водород и метан $CH_4$, а в атмосфере Юпитера есть ещё и аммиак $NH_3$.» Далее пишут, что «поскольку планеты-гиганты сильно удалены от Солнца, их температура (по крайней мере, над облаками) очень низка: на Юпитере –145°С, на Сатурне –180°С, а на Уране и Нептуне ещё ниже». Вот какими сведениями мы располагаем об атмосферах планет гигантов.

[4] Далее идёт описание Марса. Описание было изъято редактором и помещено в *Приложении* к последней работе автора «Энергетическая структура Солнца и планет солнечной системы с точки зрения механики безынертной массы» (стр. 49). Там же объяснены причины перемещения этой части текста.

Мы знаем, что атмосфера этих планет имеет идеальную шарообразную форму. Это значит, её форму образуют силы поля планетного тяготения, которым противодействуют силы теплового расширения газов атмосферы. Сама гидросфера планет-гигантов размещается в центре атмосфер этих планет. По этой причине мы можем сказать, что толщина атмосферного слоя на полюсах Юпитера и Сатурна более 6000 км, а на их экваторах толщина слоя атмосферы измеряется сотнями километров. Поскольку радиус экватора этих планет на 6000 км больше радиуса полюсов. На Уране и Нептуне над полюсами слой атмосферы равен порядка 1000 км, а на их экваторах он равен нескольким сотням километров, поскольку экваториальный радиус этих планет на 500 км больше радиуса полюса. Используя данные практических наблюдений этих планет, мы можем сказать, что атмосферы планет-гигантов многослойны и состоят из нескольких слоев газов с различной плотностью. Непосредственно у поверхности этих планет располагаются атмосферные слои с наибольшей плотностью, а у граничной поверхности атмосферы располагаются слои газов с минимальной плотностью.

Большая толщина атмосфер планет-гигантов и повышенная плотность их газов, зависящая от химического состава этих газов, требуют для поддержания своей тепловой энергонасыщенности очень большого расхода тепловой энергии. Этот тепловой расход планет-гигантов может обеспечить только гидросфера этих планет, так как они удалены от Солнца на большое расстояние. Это значит, что температура поверхности Юпитера и Сатурна должна исчисляться тысячами градусов, а Урана и Нептуна – сотнями градусов. В связи с тем, что атмосферы планет-гигантов имеют большую толщину и состоят из газов повышенной плотности, то атмосферное давление на поверхности Юпитера и Сатурна будет измеряться в тысячах атмосфер, а на поверхностях Урана и Нептуна – сотнями атмосфер. В результате того, что газы атмосфер этих планет находятся под большим давлением и осуществляют большой теплосъём с поверхностей этих планет, то поверхности планет-гигантов покрываются незначительной толщины темной корочкой. Поэтому мы не видим самостоятельного свечения этих планет. Из-за высоких температур гидросфера планет-гигантов не имеет планетной коры, как, например, Земля, то есть их гидросфера полностью находится в жидком состоянии. Это значит, что построение гидросферы планет-гигантов более схоже с построением гидросферы Солнца, чем с другими планетами.

Далее, замеренные температуры над облаками Юпитера (–145°С) и над облаками Сатурна (–180°С), говорят о том, что самый верхний слой лёгких газов атмосфер этих планет не образует критических течений и не имеет поверхности критических скоростей газов, а ограничивается сразу поверхностью конденсации. Ибо измеренные температуры соответствуют температуре конденсации более легких газов типа кислорода, азота и т. д. Подобное построение имеют атмосферы Урана и Нептуна, так как там отмечены ещё более низкие температуры. Это значит, что атмосферы этих планет состоят только из тропосферы, или подзоны энергетического прироста газов, а стратосферы, или зоны расширения газов, они не имеют. Сразу же над тропосферой будет располагаться ионосфера, или зона паров, через которую уходит тепловая энергия планет в открытый космос.

Видимые облака атмосферы планет-гигантов могут состоять только из конденсата более тяжёлых газов. Большое различие по механической энергии между экваториальными и полюсными зонами, а также высокие температуры поверхностей этих планет приводят к сильному движению газовых масс их атмосфер, которое наблюдают астрономы в свои телескопы. Вот всё то новое, что мы можем узнать о планетах-гигантах, рассматривая их с точки зрения механики безынертной массы.

### [*Строение Солнца*]

Теперь мы можем перейти к рассмотрению построения самого Солнца. Как видите, мы своё путешествие начали с разбора планет, а не Солнца. Тем самым мы провели определённую подготовку, чтобы приучить себя к непривычному. Ведь мы пытаемся посягнуть на одно из самых больших неизвестных. Что требует всегда соответствующей подготовки. Планеты остаются планетами. На одной из них мы сами живем. Поэтому мы всякие новые положения можем проверить в своих условиях обитания. Что касается Солнца, то нам вряд ли когда-либо удастся побывать там, хотя бы при помощи приборов. К тому же Солнце нам греет и светит. Как бы там ни было, но оно отличается от планет лишь повышенной температурой. Поэтому Солнце, как и всякая планета, должна иметь гидросферу и атмосферу.

Солнце вращается вокруг своей оси довольно быстро. Солнечные сутки равны двадцати пяти земным суткам. Тангенциальные скорости плоского установившегося потока гидросферы Солнца равны порядка 23000 м/сек. Это значит, что гидросфера Солнца находится в состоянии динамического равновесия. Во входном потоке плоского установившегося потока, или в центре гидросферы, располагается солнечный источник тепловой энергии. Часть тепловой энергии этого источника поглощается плоским установившимся потоком гидросферы. Поэтому температура на поверхности гидросферы несколько ниже. По температурному режиму Солнце относится к разряду жёлтых звезд. Если бы в его гидросфере не было плоского установившегося потока, то температура поверхности возросла бы, и Солнце отнесли к разряду белых звезд.

Атмосфера Солнца состоит из тропосферы и стратосферы. Тропосфера, или зона энергетического прироста газов, Солнца состоит из многих газовых слоёв различных по плотности. Ибо в атмосфере Солнца найдено 68 химических элементов периодической таблицы Д. И. Менделеева. Каждый газовый слой тропосферы ограничивается поверхностью минимальной энергии конденсации. Самый верхний слой тропосферы составляют лёгкие газы: водород и гелий. Эти газы образуют подзону постоянных температур газов, которая поверху ограничивается поверхностью критических скоростей газов. Эта поверхность является границей раздела между тропосферой и стратосферой атмосферы Солнца. В стратосфере, или в зоне расширения газов, происходит адиабатическое расширение газов до полной потери механической энергии с одновременным выделением тепловой энергии. По этой причине температура газов возрастает по мере продвижения от нижней до верхней границы стратосферы. Верхняя граница стратосферы тоже образуется поверхностью с минимальной энергией конденсации газов. По этой причине самая низкая температура атмосферы Солнца находится в подзоне постоянных температур газов. Вот так выглядит атмосфера Солнца. Выше атмосферы располагается ионосфера, через которую тепловая энергия Солнца выводится в окружающий космос.

Для зон атмосферы Солнца приняты иные обозначения отличные от наших. Согласно описанию, фотосферой называют тропосферу, или зону энергетического прироста газов, а хромосферой – стратосферу, или зону расширения газов. Будем считать, что мы привели в соответствие принятые обозначения зон атмосферы Солнца с нашими обозначениями. Следовательно, зона солнечной короны будет совпадать с ионосферой, или зоной паров.

Сделав осмотр планет и Солнца, мы получили почти одинаковую картину их строения. Теперь необходимо уточнить различия, чтобы выделить в определённые группы

На Солнце в зоне постоянных температур газа отмечена самая низкая температура 4400°К. Следовательно, к центру Солнца она увеличивается во много раз. При таких температурах любое нам известное химическое вещество может находиться в газообразном состоянии. Далее вспомним, что по нашим подсчётам тангенциальная скорость плоского установившегося потока имеет сравнительно большую величину (23000 м/сек). Плюс ко всему, средняя плотность Солнца не велика, она равна 1,4 г/см$^3$. Всё это говорит о том, что вся масса Солнца находится в газообразном и даже в плазменном состоянии. По этой причине у Солнца не должно быть разделения между гидросферой и атмосферой. Ибо оно не имеет гидросферы. Следовательно, исходя из этих выводов, мы должны будем несколько уточнить наше первое строение Солнца. Это уточнение будет заключаться в том, что ранее определённые границы для хромосферы, или стратосферы, Солнца останутся в прежних пределах, а границы фотосферы, или тропосферы, мы должны будем раздвинуть, коль мы определили, что на Солнце нет массы, находящейся в жидкой фазе.

Тогда мы должны будем наружную границу фотосферы, или тропосферы, совместить с наружной граничной поверхностью плоского установившегося потока Солнца. Тогда в новом виде Солнце будет иметь такое строение: в центре Солнца будет располагаться плоский установившийся поток насосного типа. За наружной граничной поверхностью этого потока будет располагаться фотосфера, или тропосфера, Солнца, которая состоит из многих газовых слоев, различных по объёмному весу. Наружной граничной поверхностью фотосферы является поверхность критических скоростей газов с подзоной постоянных температур газов. За поверхность Солнца мы примем поверхность критических скоростей газов. Как вы помните, эта поверхность формируется силами поля тяготения Солнца и не зависит от сил его плоского установившегося потока. Уже над этой поверхностью располагается хромосфера, или стратосфера, Солнца, которая по верху ограничивается поверхностью минимальной энергии конденсации. Выше располагается ионосфера, или корона Солнца. Вот в таком виде существует наше светило.

Планеты-гиганты: Юпитер, Сатурн, Уран и Нептун, тоже имеют большие тангенциальные скорости, порядка 6000 м/сек, в своих плоских установившихся потоках. Средняя плотность их массы по величине почти одинакова с плотностью Солнца. Отметим, что подобное сравнение будет не совсем точным. Ибо при определении средней плотности Солнца за его объём брали объём, образованный поверхностью критических скоростей газов, а при определении средней плотности планет-гигантов за объём планеты брали объём, образованный поверхностью гидросферы. Если бы объёмы этих планет взяли бы по их поверхностям критических скоростей газов, то средняя плотность оказалась бы во много раз меньше средней плотности Солнца. Для правильного сравнения плотностей планет и Солнца только так надо поступать.

Из нашего замечания напрашивается такой вывод, что основной объём планет-гигантов составляют массы, находящиеся в газообразном состоянии. Только в этом случае объясняется большое различие между средней плотностью этих планет и Солнца. Ибо в этом случае причиной подобного различия является различие уровней механической энергии Солнца с уровнем любой из этих планет. Эта причина является единственной причиной для различия средней плотности.

Тогда мы можем представить себе строение планет-гигантов в таком виде:

в центре этих планет размещается плоский установившийся поток насосного типа, рабочим телом которого является газ. Коль этот поток за свою периферийную граничную поверхность выталкивает газ, то основная масса планеты за объёмом потока тоже будет находиться в газообразном состоянии. Масса этих планет, находящаяся в жидком состоянии, будет составлять лишь незначительную часть от общей массы планеты. Ибо в этом случае она должна будет расположиться в виде планетной коры (типа земной коры) и образовать соответствующий слой определённой формы, то есть поверхность планет-гигантов образует жидкая планетная кора. Она как бы разделяет внутреннюю газообразную массу планеты с её наружной газовой массой атмосферы. Строение планет-гигантов нам известно.

Сравнивая массу планет-гигантов с массой Солнца по состоянию, мы можем отметить, что масса планет-гигантов находится в большей своей части в газообразном состоянии с промежуточным небольшим количеством массы, находящейся в жидком состоянии из-за меньшего количества тепловой энергии этих планет приходящейся на их среднюю плотность по сравнению с Солнцем. Причиной различия состояния масс являются источники тепловой энергии Солнца и планет-гигантов. Это значит, что объёмная или какая-либо другая единица источника тепловой энергии планет-гигантов излучает меньшее количество тепла, чем единица источника тепла Солнца, то есть удельная мощность теплового источника планет-гигантов меньше удельной мощности источника тепла Солнца. Поэтому мы можем сказать, что планеты-гиганты представляют собой более холодные солнца.

Планеты земной группы: Меркурий, Венера, Земля и Марс отличаются от Солнца и планет-гигантов своей большой средней плотностью. Это различие можно объяснить только тем, что основная масса этих планет находится в жидком состоянии и лишь незначительная её часть находится в газообразном состоянии в виде атмосфер этих планет.

Земля и Марс имеют в своей гидросфере плоские установившиеся потоки насосного типа. Меркурий и Венера вращаются под действием течения, которое организовано в виде плоского установившегося потока. Удельная мощность тепловыделения источников тепловой энергии этих планет меньше, чем у планет-гигантов, и во много раз меньше, чем у Солнца. По этой причине масса планет земной группы находится в жидком состоянии.

Отсюда следует, что принципиальное различие между звёздами и планетами заключается в различной удельной мощности их источников тепловой энергии. Под звёздами здесь имеется в виду только Солнце. Это значит, мы можем выделить три типа источников тепловой энергии: это солнечный источник тепловой энергии, источник энергии планет-гигантов, который схож с солнечным источником, и источник тепловой энергии планет земной группы.

Мы рассматриваем звезды и планеты с точки зрения механики безынертной массы, поэтому различие планет и звезд по химическому составу мы не можем учесть. Вот почему мы свели их основное различие к различию их источников тепловой энергии. Тем самым мы оставили свободу действий другим специалистам.

**[*Состояние изменения энергетического уровня Солнца и планет*]**

Далее мы должны будем дополнить это различие. Выше нами было установлено, что для вращающихся вокруг своей оси планет, в том числе и звёзд, существует три формы динамического равновесия. К первой форме мы отнесли планеты находящиеся просто в динамическом равновесии. Ко второй форме мы отнесли планеты находящиеся в состоянии статического равновесия с динамическим движением. Последним состоянием планеты является состояние изменения энергетического уровня. Эта форма состояния планет и звезд наступает в том случае, когда механическая энергия на выходе из плоского установившегося потока планеты превысит механическую энергию оставшейся массы планеты, которая находится под воздействием сил поля планетного тяготения. Эти условия мы записали выше. Первые два состояния мы уже рассмотрели и сделали соответствующие выводы. Теперь нам осталось рассмотреть состояние изменения энергического уровня.

Это состояние планет отличается от других их состояний, прежде всего, тем, что оно длится для них очень малое время. Ибо при изменении энергетического уровня должно происходить разрушении гидросферы планеты. Коль это состояние планет не продолжительно по времени и разрушает планеты, то оно должно оставлять следы своего пребывания на этих планетах и звёздах. Следы, как вы знаете, сохраняются очень долгое время.

В том, что Солнце находилось несколько раз в состоянии изменения энергетического уровня, свидетельствуют девять планет солнечной системы. В свою очередь, планеты, кроме Меркурия и Венеры, тоже находились в состоянии изменения энергетического уровня, о чём свидетельствуют их спутники. Например, у Юпитера их двенадцать, а у Сатурна девять спутников. Титан, самый большой спутник Сатурна, имеет атмосферу (состоящую из метана). Это значит, что при состоянии изменения энергетического уровня планеты происходит не полное её разрушение, а лишь производится отстрел части массы её гидросферы, которая затем в виде спутника соответствующей планеты или звезды продолжает существовать во времени. Здесь мы должны себе отметить, что с отстрелом части массы гидросферы происходит одновременно уменьшение общего уровня механической энергии планеты и механической энергии плоского установившегося потока планеты. После чего планета продолжает существовать, перейдя на новый энергетический уровень.

Всё это мы констатируем как факт, который подлежит дополнительному изучению. Именно как факт, так как для его количественного подтверждения мы имеем необходимые зависимости и законы, которые содержатся в механике твёрдого тела и механике безынертной массы. Но для этого ещё необходимы точные данные о планетах, спутниках и необходимое количество времени. Тогда мы сможем определить величину энергии расходуемую на отстрел спутника соответствующей планетой, звездой, и траекторию движения спутника после отстрела до занимаемой им в настоящее время орбиты. После того, как будут получены точные количественные данные по этому вопросу, то, возможно, будет установлена причина для третьего состояния планет и звезд. Поскольку у нас нет ни точных данных, ни времени, мы постараемся убедиться в правильности происхождения планет и их спутников упрощенным способом. В качестве примера рассмотрим образование планеты Земля и её спутника – Луны.

Полагаем, что в какой-то определённый период времени для Солнца наступило состояние изменения энергетического уровня. Это состояние начинается с того, что в силу возникших определённых причин начинает быстро расти величина тангенциальной скорости плоского установившегося потока Солнца. Диаметр Солнца в экваториальном положении тоже начинает расти по той причине, что энергия плоского установившегося потока начинает превосходить по величине энергию сил поля солнечного тяготения. Когда увеличение диаметра достигает определённого кризисного положения, то есть того положения, когда Солнце должно будет разрушиться, то в этот же момент должна будет отделиться часть массы Солнца. И она отделяется в виде газового облака, которое имеет высокую температуру. Далее это газовое облако начинает удаляться от Солнца. По той причине, что оно в этот период представляет собой определённый объём массы, у которого имеется центр масс, то мы обязаны будем применить к этому облаку законы механики твёрдого тела, чтобы вычислить траекторию и скорость его удаления от Солнца.

Мы рассмотрели одну версию образования газового облака. По другой версии, газовое облако может образоваться по причине перехода части энергии плоского установившегося потока в энергию акустической волны, поступательная скорость которой превышает скорость ускользания от Солнца. Мы сейчас можем утверждать только, что одна из этих версий является объяснением причины образования планет. Других причин не может быть. Ведь образование планет и спутников тоже должно идти в соответствии с законами механики безынертной массы. Причины, скажем, понуждающие массу Солнца к состоянию изменения энергетического уровня могут быть и физического, например, электромагнитного характера, и химического.

Далее газовое облако, масса которого равна сумме масс Земли и Луны, занимает свою орбиту движения вокруг Солнца в соответствии с механической энергией, которое бы имело твёрдое тело, такой же массы с такой же скоростью движения от Солнца, то есть в соответствии с законами механики твёрдого тела.

Через некоторое время после отделения от Солнца, ещё при движении к своей орбите, газовое облако начинает приобретать шаровую форму относительно своего центра массы под действием сил поля планетного тяготения. Возможно, на подлёте к орбите или уже при движении по своей орбите газовое облако приобретает идеальную шаровую форму, которая соответствует планетам и звездам, находящимся в поле сил планетного или звездного тяготения. После приобретения шаровой формы наше газовое облако ещё имело одинаковый химический состав своей массы как смесь химических элементов и веществ и равномерное одинаковое распределение энергии по всему объёму облака-планеты.

Для постороннего наблюдателя образование облака-планеты внешне выглядело бы в таком плане. Он сначала увидит очень яркую вспышку на Солнце. Ведь в этот момент разверзнутся солнечные недра и сверхвысокие солнечные температуры появятся на внешней стороне Солнца. Затем от солнечной поверхности начнёт быстро расти ослепительный газовый столб. Затем он отделится от поверхности Солнца и начнет удаляться в космическое пространство сияя ярче, чем само Солнце. Единство формы газового столба определяется образовавшимся в нем силовым полем планетного тяготения, центр которого размещается в толще газового столба. Ибо сам столб образуется как установившийся поток газа, в любом сечении которого скорость движения газов превышает скорость, необходимую для создания солнечного спутника. После отделения газового столба от Солнца эту скорость газу помогает сохранить образовавшееся в газовом столбе поле сил планетного тяготения. Оно как бы вбирает в себя кинетическую энергию массы газового столба, который после отделения от солнечной поверхности начнет существовать в своём пространственном движении аналогично твёрдому телу с постоянной скоростью, которая равна по величине скорости необходимой для образования солнечного спутника.

Далее наш наблюдатель увидел бы, как газовый столб, который мы выше назвали газовым облаком, начал постепенно терять свою продолговатую форму и приобретать округлые формы шаровой поверхности. Через некоторое время наблюдатель уже видел бы сияющий не меньше Солнца шар вместо газового столба. И мы бы стали свидетелями такой картины, когда по орбите вокруг Солнца движется маленькое солнце-планета. Вот так в наглядной форме выглядит для наблюдателя первый этап образования планеты, масса которой равна суммарной массе Земли и Луны.

Существуя на орбите, солнце-планета стало бы претерпевать следующие изменения. После того, как газовая планета приобрела шарообразную форму, в её объёме начинается расслоение газов в соответствии с их плотностью. С большей плотностью или с большим объёмным весом газы стремятся занять центр планеты и близлежащие к нему области, а газы с минимальной плотностью или минимальным объёмным весом стремятся занять периферийные области планеты. Ибо лёгкие газы вытесняются из центра планеты более тяжёлыми газами.

При вылете из объёма Солнца газовая масса планеты имела одинаковое с Солнцем количество тепловой энергии на объёмную единицу. По этой причине на первом этапе интенсивность её излучения световой и тепловой энергии была одинакова с интенсивностью излучения Солнца. В результате того, что сильная интенсивность излучения тепловой энергии в открытый космос с вновь образовавшейся планеты не пополнялась в должном количестве источником тепловой энергии этой планеты, то в последующем с течением времени началось остывание массы планеты, то есть происходило уменьшение энергонасыщенности тепловой энергией массы планеты. Для наблюдателя это выглядит как медленное угасание яркости вновь образовавшейся планеты. В какой-то момент времени она совсем перестаёт испускать свой собственный свет и начинает светиться отраженным светом Солнца. В этот период в центре планеты образуется плоский установившийся поток насосного типа. Масса планеты ещё к тому моменту находится полностью в газообразном состоянии. В связи с тем, что энергия плоского установившегося потока по каким-то причинам оказывается больше механической энергии поля планетного тяготения, для вновь образовавшейся планеты наступает состояние изменения энергетического уровня. В результате чего, эта планета отстреливает часть своей массы таким же способом, как это происходило на Солнце. От планеты отделяется часть газообразной массы, которая в последующем занимает орбиту своего движения вокруг планеты в соответствии с величиной полученной ею механической энергии. Таким же образом произошло отделение массы Луны от массы Земли. Дальнейшую судьбу Земли и Луны рассмотрим раздельно.

Земля будет продолжать своё движение по околосолнечной орбите неизменным. Ибо отстрел массы не может изменить её орбитальных характеристик. По той причине, что при отстреле массы планета находится в газообразном состоянии и центр её массы и поля тяготения совмещены. Чтобы изменить орбитальные характеристики такой планеты надо произвести одностороннее силовое воздействие либо одновременно на всю массу планеты, либо на большую её часть, а отстрел части массы вносит лишь местные силовые возмущения, которые не влияют на орбитальные характеристики планеты.

Утратив часть массы, Земля перейдет на новый уменьшенный энергетический уровень. Энергия её теплового источника тоже уменьшится, и плоский установившийся поток её гидросферы тоже уменьшит свою механическую энергию. При дальнейшем существовании Земли на своей орбите происходит балансировка её тепловой энергии в соответствии с её источником тепловой энергии. Ибо этот источник, обладая определённой количественной величиной мощности по теплопроизводительности, способен обеспечить тоже определённую тепловую энергонасыщенность массы Земли. Масса планеты после отстрела имела повышенную тепловую энергонасыщенность по сравнению с возможной энергонасыщенностью для её источника тепла, так как она находилась в газообразном состоянии. По этой причине происходило дальнейшее остывание массы Земли. Из газообразного состояния масса перешла в жидкое состояние, образовав гидросферу Земли с её газовой оболочкой, которую мы называем атмосферой. В этой стадии атмосфера Земли по своему строению и составу почти не отличалась от атмосферы Венеры.

При дальнейшем остывании гидросферы Земли начало происходить образование земной коры. Над раскаленной поверхностью гидросферы располагался углекислотный слой атмосферы. Над этим слоем размещался слой азотно-кислородной смеси атмосферы Земли. Толща водяных облаков азотно-кислородного слоя, располагаясь на верхней граничной поверхности углекислотного слоя, плотной пеленой укрывала гидросферу Земли от воздействия тепловой энергии Солнца. Образовавшийся слой земной коры, пока ещё небольшой толщины, начал уменьшать приток тепловой энергии от гидросферы Земли к её атмосфере. Что, в свою очередь, уменьшает тепловую энергонасыщенность атмосферы. Для атмосферы Земли уменьшение тепловой энергонасыщенности будет выражаться в уменьшении толщины её углекислотного слоя. Когда толщина земной коры стала достаточной, а толщина углекислотного слоя атмосферы уменьшилась, атмосферная вода в виде дождя начала попадать на поверхность Земли. Тем самым увеличилась интенсивность охлаждения земной коры. Ибо дождевая вода тут же испарялась на поверхности и возвращалась снова в атмосферу, унося с собой большое количество тепла. Поэтому толщина земной коры начала расти более быстро.

Затем началось формирование рельефа поверхности Земли. Сначала её поверхность представляла собой безбрежный океан расплавленной магмы с плавающими островами земной коры. Затем поверхность Земли покрылась сплошной тонкой коркой с полыньями жидкой расплавленной магмы. После этого на каких-то участках этой тонкой корки наращивать её толщину стали более легкие породы и минералы. Поэтому эти участки, всплывая из магмы, поднимались всё выше и выше над её поверхностью. Вот эти участки образовали современные материки нашей планеты. На других же участках тонкой коры либо её толщину наращивали более тяжёлые породы и минералы, либо медленнее шло наращивание толщины из-за сложившегося местного теплового режима. Поэтому эти участки всплыли над поверхностью магмы на меньшую высоту. В современных условиях они образовали поверхность дна морей и океанов. Так организовался рельеф поверхности нашей Земли.

После образования рельефа началось сравнительно равномерное наращивание толщины земной коры по всей её поверхности. Поэтому земная кора, как бы непрерывно всплывая по высоте над поверхностью магмы, сохранила до какой-то степени свой первоначальный поверхностный рельеф. Конечно, с течением времени поверхностный рельеф претерпевал определённые изменения. Например, на каких-то участках могла усилиться интенсивность застывания легких пород и минералов, которые под действием выталкивающих сил взламывали земную кору и образовывали горные возвышенности и цепи. Под другими участками земной коры мог усилиться теплообмен в магме. В результате чего на этих участках могло произойти уменьшение толщины земной коры, и они должны были опуститься ниже первоначальной своей высоты и т.д. В современных условиях земная кора тоже испытывает непрерывные изменения, но общая её масса и средняя её толщина остаются неизменными. Ибо эти величины диктуются тепловой производительностью источника тепла нашей планеты, которая определяет тепловую энергонасыщенность массы Земли.

Вернемся немного назад и рассмотрим развитие жизненных условий на Земле. Когда сформировался рельеф земной поверхности, но он возвышался над уровнем магмы на небольшой высоте, температура на поверхности материков достигала сотни градусов, а на поверхности будущих океанов она всё ещё достигала нескольких сотен градусов. Поэтому на материковых поверхностях вода начала сохранятся в кипящем состоянии, а с поверхностей океанов она испарялась. В этот период в атмосфере Земли существовала мощная водная циркуляция. Над материковыми поверхностями вода выпадала в виде мощных дождевых потоков, и на поверхности дна будущих океанов она превращалась в пар и возвращалась снова в атмосферу.

Наконец, когда толщина земной коры возросла настолько, что вода в океанах стала сохраняться в кипящем состоянии, а в бассейнах материков она имела температуру несколько десятков градусов, солнечный свет стал проникать на поверхность Земли. Будем считать, что этот период стал началом зарождения органической жизни на Земле. В теплых бассейнах материковой воды произошло рождение органической жизни Земли. Мы утверждаем, что органическая жизнь на Земле произошла на её материках потому, что и в наши дни в какой-то степени ещё сохранилось различие в материковом разнообразии жизненных форм живых организмов. Например, на одних материках мы можем встретить отдельные виды животных, которых нет на других материках.

Затем толща земной коры возросла настолько, что температура на поверхностях материков и океанов стала сравнительно одинаковой и достигла нескольких десятков градусов. В этот период на всей поверхности Земли образовался тёплый и очень влажный климат. Тепловой режим поверхности и атмосферы Земли поддерживался в основном за счёт её внутреннего тепла. Солнечная энергия в этих условиях необходима была живым организмам только как источник световой энергии. Времена года определяли лишь продолжительность освещённости полушарий Земли. На Земле этого периода, если так можно выразиться, был парниковый климат. Вот это парниковый климат вызвал к жизни гигантские формы растений и животных, в том числе насекомых. Для поедания крупной и буйно растущей растительности нужны были крупные травоядные животные, а для поедания крупных травоядных нужны были крупные хищники[5]. Теплолюбивые растения росли в это время и на теперешней Антарктиде. Вода на Земле находилась либо в жидком, либо в парообразном состоянии. Льда и снега не было. Для животных парникового и очень влажного климата были не приемлемы шерсть и перья современных животных. По этой причине даже летающие ящеры имели перепончатые крылья.

Наконец, когда толща земной коры стала настолько большой, что почти полностью отграничила атмосферу Земли от её внутреннего тепла, климат на Земле резко изменился. Он стал более сухим и более холодным. Преобладающее значение для теплового баланса стала иметь тепловая энергия Солнца, и лишь незначительную роль играет тепловая энергия гидросферы Земли. В общем, климат и жизненные условия на поверхности Земли стали более суровыми. Поэтому растения и живые организмы, приспособленные к условиям парникового климата, быстро вымерли и их место заняли другие живые организмы, приспособленные к современным жизненным условиям. Для северной Европы изменение климата Земли охарактеризовалось ледниковым периодом. В области полюсов Земли образовались вечные льды. В общем, для массы Земли наступила полная стабилизация по тепловой энергонасыщенности относительно её источника тепла. Отступление ледника с северной Европы можно объяснить лишь образованием теплого течения Гольфстрим в Атлантическом океане[6]. Мы теперь знаем, что движение и состояние жидкости, в том числе воды, определяется силовым полем. Поэтому существование теплых и холодных течений океанов мы можем объяснить только наличием в этих зонах локальных силовых полей, природа которых нам ещё неизвестна. В настоящее время уже известны не только океанические течения, но и их противотечения, то есть течения, которые расположены ниже поверхности течения и имеющие противоположно направленное движение воды поверхностному течению. Поэтому мы должны признать, что местные силовые поля гидросферы Земли играют немаловажную роль для её биологической жизни. Вот все те особенности, которые мы можем определить с точки зрения механики безынертной массы для нашей планеты.

Теперь проследим дальнейшую судьбу Луны. Луна, отделившись от Земли и получив от неё соответствующее количество механической энергии, аналогично твёрдому телу займет околоземную орбиту и продолжит своё движение. Одновременно она получает от Земли определённую тепловую энергонасыщенность. Ибо масса Луны в этот период находится в газообразном состоянии. Мы теперь

---

[5] в т.ч. крупные падальщики.

[6] автор предполагал, что температура на земной поверхности может также зависеть от состояния системы циркуляции подземных вод, см. «Энергетическая структура Солнца и планет солнечной системы с точки зрения механики безынертной массы». Ледниковых периодов было несколько, поэтому не исключено, что движение коры сильно влияло на эту систему, то разрушая, то давая время ей восстановиться.

Кроме того, под полюсами Земли внутреннее вещество Земли имеет более низкую температуру, поскольку вещество подходит к полюсам остывшим. Пока неизвестно, имеет ли внутренний тепловой источник постоянную мощность или переменную. Как известно, на Солнце бывают периоды местного охлаждения в виде темных пятен и выбросы вещества – протуберанцы. Что говорит о переменной мощности внутреннего источника тепла. Автор высказал самые общие соображения об истории климата Земли. Мы должны понимать, что история климата, причины его изменений во многом обусловлены внутренним строением Земли и не могут быть объяснены без исследования этого внутреннего строения.

знаем, что поток тепловой энергии нельзя остановить и превратить в застывшее движение. Поток тепловой энергии всегда является действительным потоком, который непрерывно движется. Поэтому газообразная масса Луны будет непрерывно излучать тепловую энергию в открытый космос. Через некоторое время в результате потери тепловой энергии масса Луны перейдет из газообразного состояния в жидкое. Образуется гидросфера Луны и её атмосфера из соответствующих газов. В гидросфере и атмосфере Луны жидкости и газы тоже распределяются послойно в зависимости от величины плотности.

Дальнейший отток тепловой энергии в открытый космос приведёт к образованию лунной поверхностной коры. Как нам известно, стабилизирующей основой для оттока тепла является внутренний источник тепловой энергии планеты или спутника, который определяет тепловую энергонасыщенность массы. Коль такого источника тепловой энергии нет внутри спутника, то отток тепловой энергии наглядно выражается как непрерывный рост толщины лунной коры. Поэтому лунная кора будет непрерывно расти по толщине, подвигаясь всё ближе и ближе к центру. В свою очередь, атмосфера Луны будет двигаться за убывающей границей гидросферы, тоже ближе к её центру. Ведь атмосфера любой планеты порождена застывшим движением теплового расширения её массы. Это значит, что атмосфера Луны движется через щели, разломы и прочие пустоты и рыхлоты в лунной коре к её центру за отступающей гидросферой. В конечном итоге лунная атмосфера должна будет поглотиться застывшей твёрдой массой Луны. По этой причине отсутствие атмосферы на Луне говорит о том, что Луна не имеет внутреннего источника тепла, а потому представляет собой застывшую твёрдую глыбу лунной массы. Возможно, у её центра ещё есть какое-то небольшое количество жидкой расплавленной массы, но это вопрос лишь времени. Коль Луна не имеет своего источника тепла, то рано или поздно этот остаток жидкой лунной массы превратится в её твёрдую массу. По этой причине не только Луну, но и другие подобные ей планеты и спутники, которые не имеют своей атмосферы, можно считать твёрдым остывшим телом. О том, что масса Луны когда-то была в жидком состоянии, говорит хотя бы её идеальная шарообразная форма. Лунные цирки и её горные цепи и им подобные факты говорят о том, что жидкая лунная масса когда-то остывала. Так что все эти факты лишний раз подтверждают правильность наших выводов.

Мы рассмотрели образование Земли и её спутника Луны. Все остальные планеты солнечной системы произошли аналогичным способом. Девять планет солнечной системы свидетельствуют о том, что Солнце девять раз находилось в состоянии изменения энергетического уровня[7]. Это значит, что все планеты солнечной системы являются частью самого Солнца. Поэтому их можно назвать спутниками Солнца. В свою очередь большинство планет тоже имеют свои спутники. Это значит, что в какое-то время они тоже находились в состоянии изменения энергетического уровня. У Юпитера, например, двенадцать спутников. Следовательно, он находился в состоянии изменения энергетического уровня двенадцать раз. У Сатурна десять спутников, плюс ко всему он имеет ещё кольца. Кольца Сатурна тоже должны быть результатом его состояния изменения энергетического уровня. Только в этом случае Сатурн произвел выброс части массы в жидком состоянии по своему экваториальному диаметру. Вот, собственно, всё, что мы можем сказать о Солнце и планетах солнечной системы.

Отметим, что, следуя общей тенденции в образовании планет и спутников, мы можем выделить особенность в образовании внутренних источником тепловой энергии этих планет и спутников. Поскольку мы не знаем ещё точного химического и физического состава спутников и планет, то отметим как факт, что наличие источника тепловой энергии для отделившейся массы зависит от её первоначальной тепловой энергонасыщенности. Планеты, которые образовались от Солнца, имели предельно возможную по максимуму тепловую энергонасыщенность своей массы, и все они имеют свой внутренний источник тепловой энергии, если даже их масса меньше массы отдельных спутников планет. Все спутники планет, за исключением самого большого спутника Сатурна, название которому Титан, не имеют внутреннего источника тепловой энергии. Надо полагать, что отстрел этого спутника произошел в тот момент, когда масса Сатурна имела тепловую энергонасыщенность близкую к солнечной, то есть отстрел массы Титана Сатурн должен был произвести почти сразу после того, как он сам отделился от Солнца.[8]

---

[7] Если учитывать пояс астероидов, который образовался в результате разрушения планеты, то изменение энергетического уровня Солнца происходило 10 раз. Известно, что звёзды могут разрушаться, теперь можно предположить, что разрушаться могут и планеты. См. «Энергетическая структура Солнца и планет солнечной системы с точки зрения механики безынертной массы».

[8] Насколько редактор помнит, автор предполагал, что кольцо Сатурна образовалось на более поздних сроках эволюции планеты, когда её вещество стало более плотным. Избыток вещества не отстреливался, а представлял собой струю.

Далее отметим, что перспективной планетой, где в каком-то будущем может быть организована органическая жизнь подобно земной, является планета Венера. Ибо все основные показатели очень близки к земным. Для создания необходимых условий для органической жизни на этой планете от человечества потребуется всего-навсего создать внутри планеты плоский установившийся поток насосного типа, поскольку источник тепловой энергии внутри этой планеты уже есть. Возможно, в этих целях от человечества потребуется ввести в массу Венеры несколько астероидов, чтобы создать необходимое энергетическое возмущение, которое способствовало бы возникновению плоского установившегося потока[9]. Возможно, это будут какие-то другие средства, не следует сейчас гадать. Всё это требует соответствующих знаний и подготовки.

О Марсе мы можем сказать, что это наиболее удобная планета для исследования и изучения энергетических режимов связанных с плоским установившемся потоком и источником тепловой энергии Марса. Ибо в любых местах поверхности Марса можно разместить автоматические действующие станции. Ведь поверхность его сухая и толщина планетной коры не так уж велика. Все остальные планеты можно будет изучать практически лишь по их спутникам. Ибо температуры их очень высоки. В атмосфере планет-гигантов можно будет размещать лишь плавающие автоматические станции.

<…>Возьмём другой пример из нашей работы. Рассматривая атмосферы планет и Солнца, мы открыли такое ранее неизвестное явление природы, как трансформацию температур в атмосфере. Ибо в настоящее время повышенные температуры в зоне стратосферы атмосфер планет объясняют тем, что в этой зоне разогрев газа идёт за счёт ультрафиолетового излучения солнечной энергии, которую, якобы, в этой зоне поглощают газы. Тогда, спрашивается, за счёт чего повышается температура газа в стратосфере Солнца? Здесь мы можем претендовать на открытие ранее неизвестного явления природы лишь потому, что для тепловой энергии ещё не найдены присущие ей законы природы. Как видите, мы не задавались целью сделать подобное открытие. Мы его сделали просто случайно, стараясь понять физическую сущность атмосферы. То есть мы изучали одно явление природы и случайно заметили другое явление природы, коль мы теперь знаем, что представляет собой трансформация температуры. Просто мы отметим, что подобным открытием мы лишь подготовили почву для действительного открытия законов природы тепловой энергии. Ведь мы тем самым установили, что тепловая энергия не может существовать в природе как застывшее движение. Она всегда, во всех случаях реализуется только как реальное, или действительное, тепловое движение. Это положение мы зафиксировали понятием тепловой энергонасыщенности массы. Подобное поведение тепловой энергии было замечено ранее при получении такой термодинамической величины, как энтропия. После появления этой величины у учёных появилось два мнения: одни говорили, что нашу планету ожидает тепловая смерть, коль тепловая энергия из неё уходит, согласно показаниям энтропии. Другие учёные им возражали, что такого не может быть. Первые учёные основывали своё мнение на реальных фактах, вторые – на простой своей убежденности в том, что коль тепловая энергия не исчезала на Земле, даже до их пребывания, то она не исчезнет никогда. Поэтому они с помощью определённой игры слов доказали, что Землю не ожидает тепловая смерть. Мы же по этому вопросу можем высказать вполне определённое мнение. Оно будет звучать так, что Землю в конечном итоге действительно ожидает тепловая смерть. Ибо Земля обязана наличием своей атмосферы только своему внутреннему источнику тепла. Как только он иссякнет, исчезнет атмосфера Земли, так как она является результатом застывшего теплового расширения газовой массы Земли. И на Земле образуются такие же условия, которые мы сейчас имеем на Луне. Следовательно, тепловая смерть Земли находится в прямой зависимости от состояния её источника тепловой энергии.[10] Таким способом мы примирили два различных мнения ученых – оптимистов и пессимистов - по этому вопросу и тем самым улучшили, или подготовили, почву для открытия в недалеком будущем законов природы, относящихся к состоянию тепловой энергии. Ибо лжеоткрытия[11] неизвестных явлений

---

[9] Автор полагал, что плоское установившееся течение недр Венеры можно преобразовать в плоский установившийся поток насосного типа. В таком случае температура на поверхности снизится, атмосфера изменится. См. «Энергетическая структура Солнца с точки зрения механики безынертной массы» (1977).

[10] В этой работе автор написал, что источником энергии является термоядерная реакция, но по многим причинам он был с этим не согласен, в том числе потому, что небесные тела имеют очень разный химический состав. В работе «Энергетическая структура Солнца и планет солнечной системы с точки зрения механики безынерной массы» изложены выводы, к которым он пришёл.

[11] В смысле, что обнаружение новых явлений, свойств, фактов не является открытием в собственном смысле этого слова. Обнаружение не является шагом вперёд, оно готовит шаг, подсказывает направление, помогает поискам. Только открытие является развитием науки, шагом вперёд. Науку двигают открытия. Если подойти с этой меркой к науке, то мы увидим, что она давно стоит на месте, в тупике и без признаков жизни. Мы вообще перестали понимать, что такое наука. Поэтому автор, который неожиданно для себя столкнулся с подобным непониманием, посчитал нужным напомнить.

природы носит случайный характер, а открытие законов природы, или действительные открытия, готовятся трудом многих поколений людей, а завершить этот труд удаётся лишь некоторым из них в своё время. Ведь в настоящее время накоплено более чем достаточное количество фактов для открытия многих ещё неизвестных законов природы. <…>

Вот мы и завершили своё путешествие по солнечной системе. Вы, наверное, подумаете, что мне о планетах всё было известно заранее и мне ничего не оставалось, как сесть и написать всё это. Нет, всё это не так просто, как вы думаете. Мне были известны лишь законы и положения механики безынертной массы, которые мною были написаны ранее, и общеизвестные положения о планетах и звездах. Вот и всё. Всё новое о планетах мне становилось известным по мере того, насколько продвигался письменный текст этой работы. Так что мне тоже было интересно узнать, что представляют собой планеты и Солнце, а без написанного я этого не смог бы сделать. Следовательно, письменный текст обязателен для познания нового. <…>

***

По всем вопросам можно обращаться к редактору по адресу: arodionova@rambler.ru

август – сентябрь 2022г.